\documentclass[fleqn,usenatbib]{mnras}
\usepackage{newtxtext,newtxmath}

\usepackage[T1]{fontenc}
\usepackage{ae,aecompl}
\newcommand{\MEarth}{\rm M_{\oplus}}
\newcommand{\Mp}{M_{\rm p}}
\newcommand{\Rp}{R_{\rm p}}

\newcommand{\pderiv}[2]{\ensuremath{\frac{\partial #1}{\partial #2}}}
\newcommand{\vpl}{v_{\rm pl}}
\newcommand{\vdust}{v_{\rm dust}}
\newcommand{\St}{\rm St}
\newcommand{\Msolar}{M_\odot}

\usepackage{graphicx}	
\usepackage{amsmath}	
\usepackage{amssymb}	
\usepackage{cancel}
\usepackage[normalem]{ulem}
\usepackage{breqn}
\usepackage[percent]{overpic}
\usepackage{xcolor}

\usepackage{etoolbox}
\makeatletter
\patchcmd\@combinedblfloats{\box\@outputbox}{\unvbox\@outputbox}{}{\errmessage{\noexpand patch failed}}
\makeatother

\definecolor{kellygreen}{rgb}{0.3, 0.73, 0.09}
\newcommand{\green}[1]{\textcolor{red}    {#1}}

\title[Dust rings formed by migrating planets]{Is the ring inside or outside the planet?:  The effect of planet migration on dust rings}

\author[Meru et al.]{
\parbox{\textwidth}{Farzana Meru,$^{1,2,3}$\thanks{E-mail: f.meru@warwick.ac.uk}
Giovanni P. Rosotti$^{3}$,
Richard A. Booth$^{3}$,
Pooneh Nazari$^{3,4}$,
and Cathie J. Clarke$^{3}$\vspace{0.5cm}}
\\
$^{1}$Department of Physics, University of Warwick, Gibbet Hill Road, Coventry, CV4 7AL, UK\\
$^{2}$Centre for Exoplanets and Habitability, University of Warwick, Gibbet Hill Road, Coventry CV4 7AL, UK\\
$^{3}$Institute of Astronomy, University of Cambridge, Madingley Road, Cambridge, CB3 0HA, UK\\
$^{4}$SUPA, School of Physics \& Astronomy, University of St Andrews, North Haugh, St Andrews, Scotland, KY16 9SS, UK\\
}

\date{Accepted XXX. Received YYY; in original form ZZZ}

\pubyear{2018}

\begin{document}
\label{firstpage}
\pagerange{\pageref{firstpage}--\pageref{lastpage}}
 \maketitle

\begin{abstract}

Planet migration in protoplanetary discs plays an important role in the longer term evolution of planetary systems, yet we currently have no direct observational test to determine if a planet is migrating in its gaseous disc.  We explore the formation and evolution of dust rings -- now commonly observed in protoplanetary discs by ALMA -- in the presence of relatively low mass (12-60 $\MEarth$) migrating planets.  Through two dimensional hydrodynamical simulations using gas and dust we find that the importance of perturbations in the pressure profile interior and exterior to the planet varies for different particle sizes.  For small sizes a dust enhancement occurs interior to the planet, whereas it is exterior to it for large particles.  The transition between these two behaviours happens when the dust drift velocity is comparable to the planet migration velocity. We predict that an observational signature of a migrating planet consists of a significant outwards shift of an observed midplane dust ring as the wavelength is increased.


\end{abstract}

\begin{keywords}
planets and satellites: dynamical evolution and stability -- planets and satellites: formation -- planet- disc interactions -- protoplanetary discs -- methods: numerical
\end{keywords}



\section{Introduction}

Our knowledge of proto-planetary discs is being revolutionised by new observational facilities like the Atacama Large Millimeter Array (ALMA), the Spectro- Polarimetric High-contrast Exoplanet REsearch (SPHERE) instrument on the Very Large Telescope (VLT) and the Gemini Planet Instrument (GPI) on the Gemini Telescope. These facilities are revealing that most discs, when observed at high resolution, show conspicuous sub-structures at small spatial scales, such as spirals \citep[e.g.][]{MWC758_spirals_cheat,HD100453_spirals}, crescents (e.g., \citealt{Nienke_dust_trap}) and rings and gaps \citep[e.g.,][]{ALMA_HLTau_cheat,TWHya_gap_GPI}.

It is very tempting to link these structures to the presence of forming planets in their discs. Indeed, it is well known that planets perturb their natal discs through their gravitational influence \citep{Goldreich_Tremaine1980}. Depending on the planet mass and the disc conditions, these perturbations have different morphologies explaining, at least in principle, the variety found in observations.  These structures can therefore be used as signposts of planets, making it possible to reconstruct the properties of the unseen planet. In turn, by comparing the results with theoretical expectations we can improve our understanding of planet formation. This information is particularly precious given that, while the detection of exoplanets around main sequence stars has been highly successful (thousands of exoplanets are now known\footnote{See the NASA Exoplanet Archive at \url{https://exoplanetarchive.ipac.caltech.edu/}. While the majority of the planets found by Kepler are not confirmed, the false positive probability is estimated to be only 5-10 percent \citep{Fressin2013,Morton_Johnson2011}}), only a handful of exoplanets have been detected around pre-main sequence stars \citep{vanEyken_etal2012_cheat,Ciardi_etal2015,David_etal2016,Donati_etal2016,Johns-Krull_etal2016,Biddle_etal2018}. We also note that alternative mechanisms for spirals (such as gravitational instabilities e.g. \citealp{Kratter_Lodato_GI}; and shadows e.g. \citealp{Montesinos_shadows_spirals}), crescents (such as vortices e.g. \citealp{Regaly_deadzone_vortex,Ataiee_ecc_planet_vortex}; and eccentric planets e.g. \citealp{Ataiee_ecc_planet_vortex}) rings and gaps (such as dead zone edges e.g. \citealp{Ruge_etal2016,Pinilla_deadzones}; dust processes e.g. \citealp{Birnstiel_dust_evolution_gap,Okuzumi_sintering_rings,Pinilla_ice_lines,Dullemond_viscous_ring_instability}; and photoevaporation e.g. \citealp{Alexander_PPVI,Ercolano_TWHya_photoevaporation}) have been proposed.  We will not contribute to this particular debate, but note that planet formation is an extremely frequent phenomenon (at least 50 per cent of stars have planetary systems, \citealt{Mayor_HARPS_occurence_freq,Fressin2013}). Therefore, while not all observed structures are necessarily the signposts of planets, it is not inconceivable that at least a subset of the observed structures are due to planets interacting with the protoplanetary disc.

One of the biggest unknowns in planet formation is the role of disc migration. Due to the limited mass available in the inner disc, disc driven migration is a crucial ingredient of planet formation models \citep[e.g.][]{Alibert_etal2006} that assemble planetary cores outside the water snowline and bring them closer to the star (in contrast to the opposite view of \textit{in-situ} formation, (e.g. \citealt{Chiang_Laughlin2013}). Migration is for example invoked in the context of the Solar System to explain the asteroid belt and the small mass of Mars (Grand Tack scenario, \citealt{GrandTack}). While migration is an old topic \citep{Goldreich_Tremaine1980}, it is far from being well understood (see \citealt{Kley_migration_review,Baruteau_PPVI} for recent reviews). The least understood regime is the so-called Type I migration, valid for planets up to roughly Neptune mass. In this range of planet masses migration is calculated to be fast (with typical timescales of $\approx \rm O(100)$ orbits), yet very sensitive to the disc conditions close to the planet \citep{Paardekooper_TypeI_torque_diffusion,Bitsch_stellar_irradiation_outward_mig}, making theoretical predictions a challenge: even the direction of migration is unclear.

This issue calls for observational benchmarks, which are currently lacking. Since we only see the end product of planet formation around main sequence stars, migration cannot be disentangled from the other phenomena affecting planet formation. For example, while hot Jupiters might be seen as evidence of disc migration\footnote{We know however of at least one case (the star CI Tau) in which the hot Jupiter is likely to have been produced via disc migration since the system still possesses a proto-planetary disc \citep{Johns-Krull_etal2016,Rosotti_eccentric_HJ_CITau}.} \citep{Lin_etal1996}, they could also be produced by dynamical scattering after the disc has dispersed \citep{Rasio_Ford1996}. As another example, while the abundance of planets observed near mean motion resonances in the Kepler multi-planet population \citep{Lissauer_Kepler_cheat} is suggestive of migration \citep{Lee_Peale_MMR_migration}, it could also be explained by tidal interactions between planets \citep{Lithwick_Wu2012}.

In this paper we investigate the observational signatures of planets undergoing migration in proto-planetary discs. We focus on axisymmetric structures, i.e. gaps and rings, now commonly observed by ALMA. The most well known example is the disc around the young star HL Tau \citep{ALMA_HLTau_cheat}, but ALMA is revealing that many other discs show rings: HD163296 \citep{Isella_rings_HD163296_ALMA}, Tw Hya \citep{Andrews_TWHydra}, HD97048 \citep{vanderPlas_cavity_substructures_HD97048}, HD169142 \citep{Fedele_rings_gaps_HD169142_ALMA}, AS 209 \citep{Fedele_AS209_gaps_ALMA}, Elias 24 \citep{Dipierro_rings_gaps_Elias24_ALMA}.

A number of authors have shown, starting from the seminal work of \citet{Paardekooper2004_dust}, that relatively low mass planets can open gaps and rings in the dust, and are observable with ALMA.  This is in contrast to other structures such as spirals and vortices that require planets with masses comparable or even higher than Jupiter \citep{Juhasz_spirals_scattered_light,Dong_spirals_scattered_light,deValBorro_vortex_GP}. In particular, \citet{Rosotti_min_detectable_Mp} quantified the threshold for gap opening in the dust to be $\approx 10 M_\oplus$ for typical disc parameters (see also the dust gap opening criterion of \citealt{Dipierro_dust_gap_criterion}).  However, planet migration has either not been included in these models, such that a stationary planet on a fixed circular orbit has been assumed, or the simulations have not involved planets migrating to a significant extent.  Migration is particularly important since planets that have not opened a gap in the gas are believed to undergo fast Type I migration.

In this paper we study how planet migration influences the formation of gaps and rings, and whether we can use the recent observations previously discussed to put observational constraints on planetary migration. We shall see that migrating planets create rings inside or outside their orbit depending on the dust size.  Since the emissivity of dust grains depends on the maximum grain size in the mixture, this in principle means that multi-wavelength imaging could reveal the differential concentration of grains of different sizes in different structures, thus offering an observational test to assess whether a planet is migrating.

In Section~\ref{sec:method} we describe our method and simulation setup.  In Section~\ref{sec:results} we show the effect of migration on the gas and dust separately.  We discuss our findings and conclude in Sections~\ref{sec:disc} and~\ref{sec:conc}, respectively.





\section{Method}
\label{sec:method}

\subsection{Numerical method}

We run 2D multi fluid simulations where we evolve the dust and gas at the same time using the \textsc{fargo-3d} code \citep{fargo3d}. The algorithm is well tested and has been used many times for protoplanetary disc studies (see \citealt{deVal-Borro_migration_comparison_cheat} for an algorithmic comparison).  The code solves the mass, momentum and energy equations of hydrodynamics:

\begin{equation}
\pderiv{\Sigma}{t} + \nabla \cdot (\Sigma \mathbf{u})
\end{equation}

\begin{equation}
\pderiv{\mathbf{u}}{t} + \mathbf{u} \cdot \nabla \mathbf{u} + \frac{\nabla p}{\Sigma} = \mathbf{g}
\end{equation}
and

\begin{equation}
\pderiv{e}{t} + \mathbf{u} \cdot \nabla e + \frac{p}{\Sigma} \nabla \cdot \mathbf{u} = 0,
\end{equation}
where $\Sigma$ is the disc surface mass density, $\mathbf u$, $e$ and $p$ are the gas velocity, specific internal energy and pressure, and $\mathbf{g}$ is the gravity term.

As described in detail in \citet{Rosotti_min_detectable_Mp}, the \textsc{fargo-3d} gas hydrodynamics code has been extended to include dust, treating the dust as a pressureless fluid that is coupled to the gas via linear drag forces.  The equation for the dust velocity, $\mathbf{v_d}$, is given by

\begin{equation}
\frac{{\rm d} \mathbf{v_d}}{{\rm d}t} + \mathbf{v_d} \cdot \nabla \mathbf{v_d} = - \frac{1}{t_s} ( \mathbf{v_d} - \mathbf{u} (t)) + \mathbf{a_d}
\end{equation}
where $\mathbf{a_d}$ are the non-drag accelerations felt by the dust and $t_s$ is the stopping time.  We employ a semi-implicit time integration algorithm for the dust, which has the advantage that it does not require short timesteps even when dealing with tightly coupled dust. These approximations are valid for low dust-to-gas ratios and particles with Stokes number $\mathrm{St} \lesssim 1 $ \citep{Garaud_etal2004}.  The effect of dust on the gas has been neglected.

We use 2D cylindrical coordinates with 1024 and 440 uniformly-spaced cells in the azimuthal and radial directions, respectively.  With this resolution the planet's Hill radius at its initial location is resolved with $\approx 4-6$ grid cells.  We also perform a resolution test whereby we double the number of grid cells in each direction to show that the main result presented in this paper still holds (Appendix~\ref{sec:appendixC}).  We use dimensionless units in which the orbital radius of the planet ($R_{\rm p}$) is initially at unity and the unit of mass is that of the central star, while the unit of time is the inverse of the Keplerian frequency of the planet at $\Rp = 1$.  We use non reflecting boundary conditions at both boundaries for the gas.  While \cite{Rosotti_min_detectable_Mp} fixed the dust density to its initial value at the inner boundary, here we allow the dust density to drop below this. To this end, we use a power-law extrapolation of the dust density at the inner boundary, limiting its value to be no more than the initial value to avoid generating arbitrary amounts of dust at the inner boundary.




Along with the gas dynamics we integrate the evolution of the dust in response to the gas disc structure. To keep our simulations scale-free we fix the Stokes number, $\rm St$, of the grains rather than their physical size. We use five dust sizes with logarithmically spaced Stokes numbers: $2 \times 10^{-3}$, $6 \times 10^{-3}$, $2 \times 10^{-2}$, $6 \times 10^{-2}$ and 0.2.

\subsection{Simulation setup}

\begin{table}
\centering
  {\small
    \begin{tabular}{lll}
    \hline
    $\Mp$ [$\MEarth$] & $\tau_I$ [orbits] & $\tau_{I, \rm lin}$ [orbits]\\
    &  (measured from simulations) & (analytical)\\
   \hline
    \hline
    12 & 2,064 & 2,653\\
    20 & 1,333 & 1,592\\
    30 & 1,118 & 1,061\\
    60 & 327 & 531\\
   \hline
  \end{tabular}
}
  \caption{Table showing the migration timescale, $\tau_I$, as measured from the simulations and the analytical values given by equation~\ref{eq:tauI_lin}.  The measured values are within a factor of 2 of the analytical estimates.}
 \label{tab:tau_I}
\end{table}

\begin{figure*}
\includegraphics[width=1.0\columnwidth]{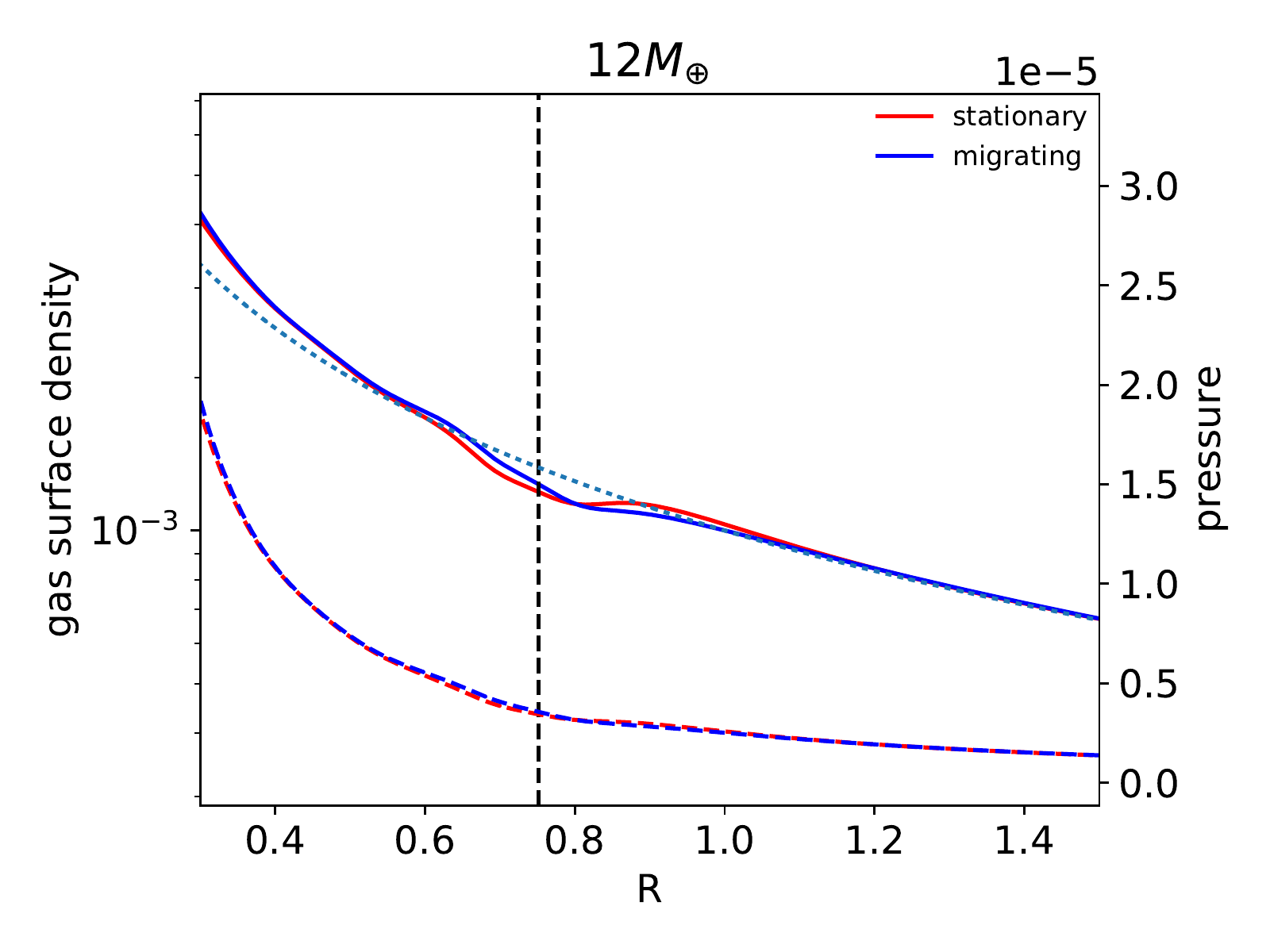}
\includegraphics[width=1.0\columnwidth]{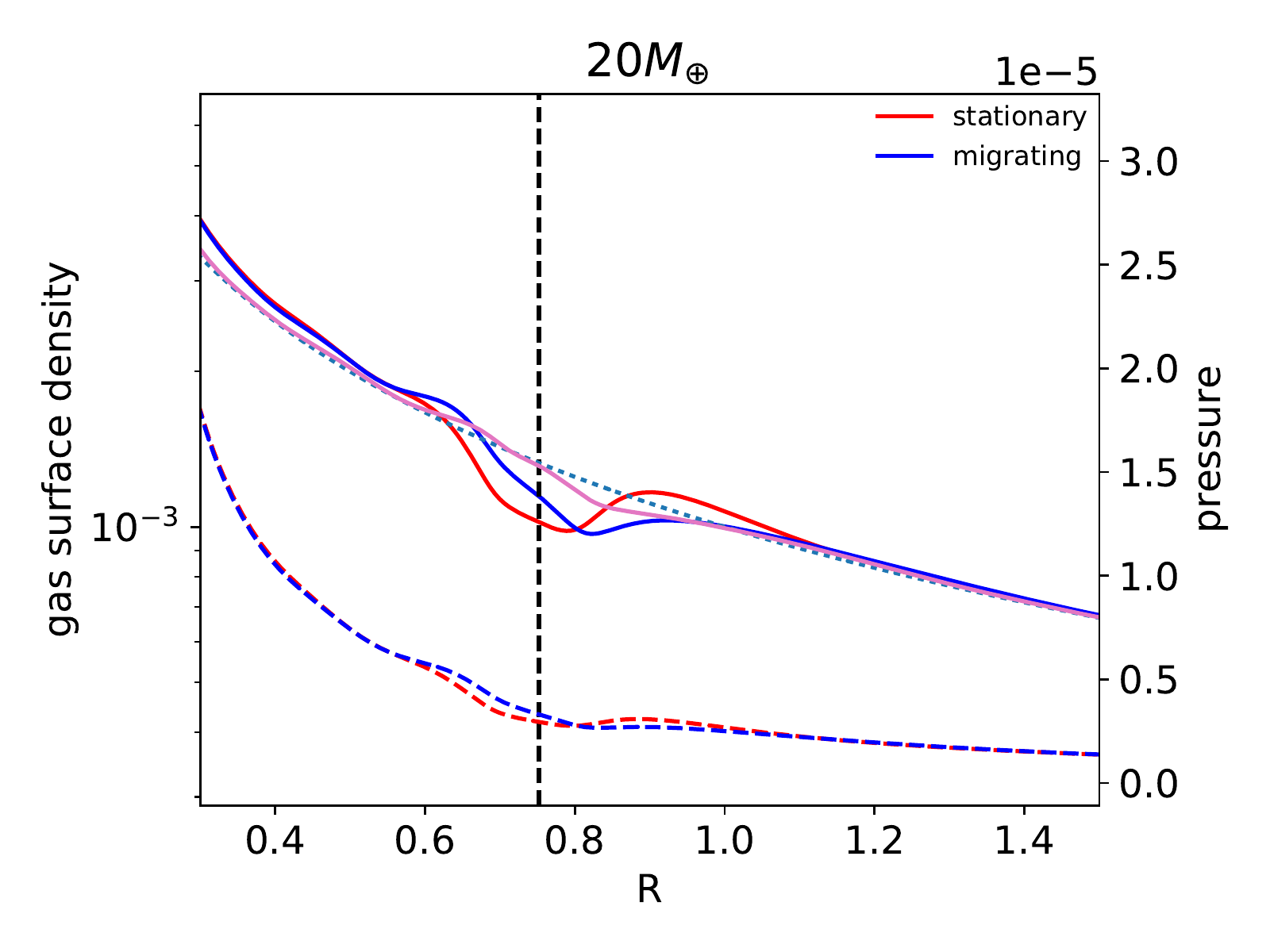}
\includegraphics[width=1.0\columnwidth]{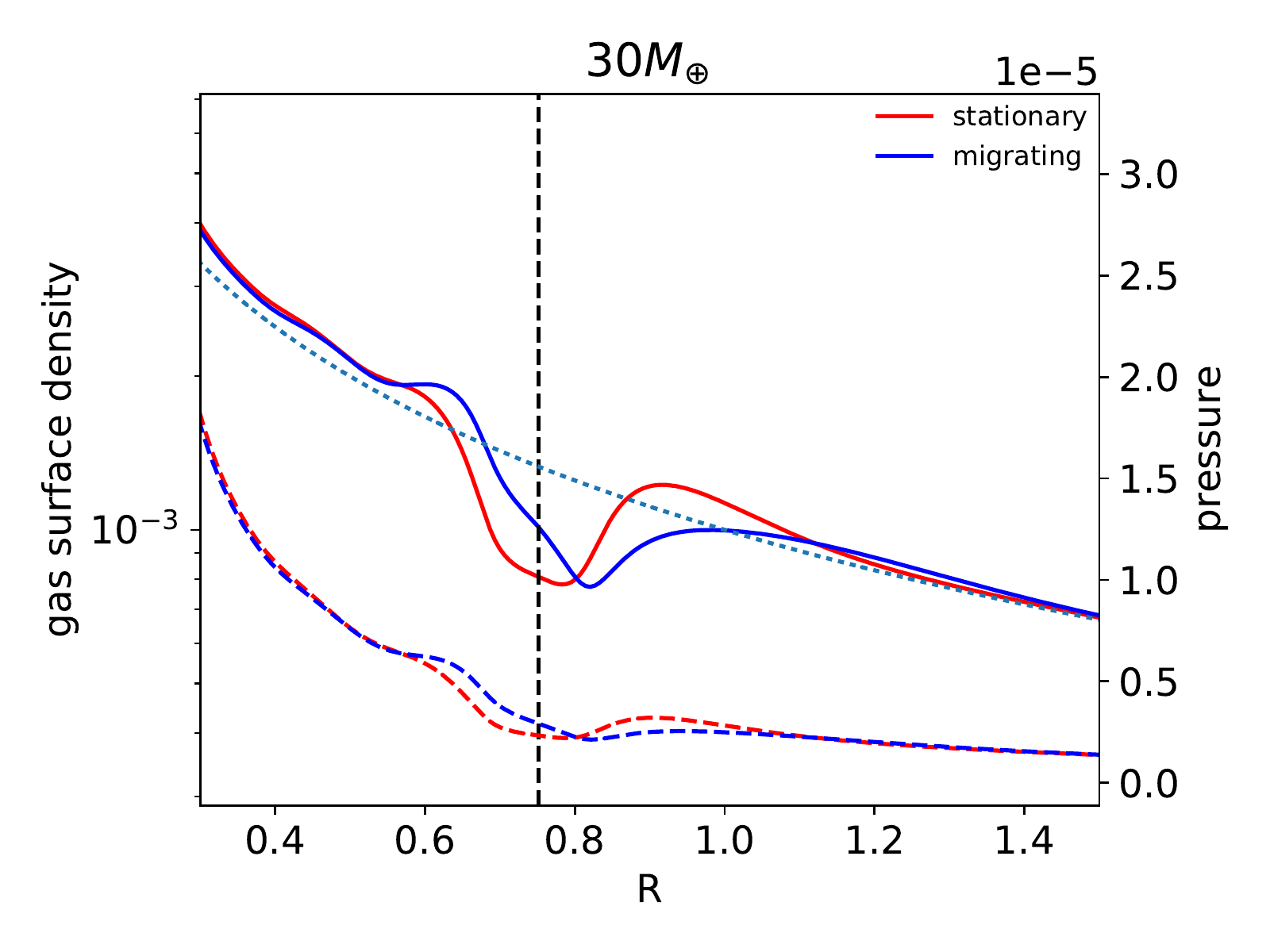}
\includegraphics[width=1.0\columnwidth]{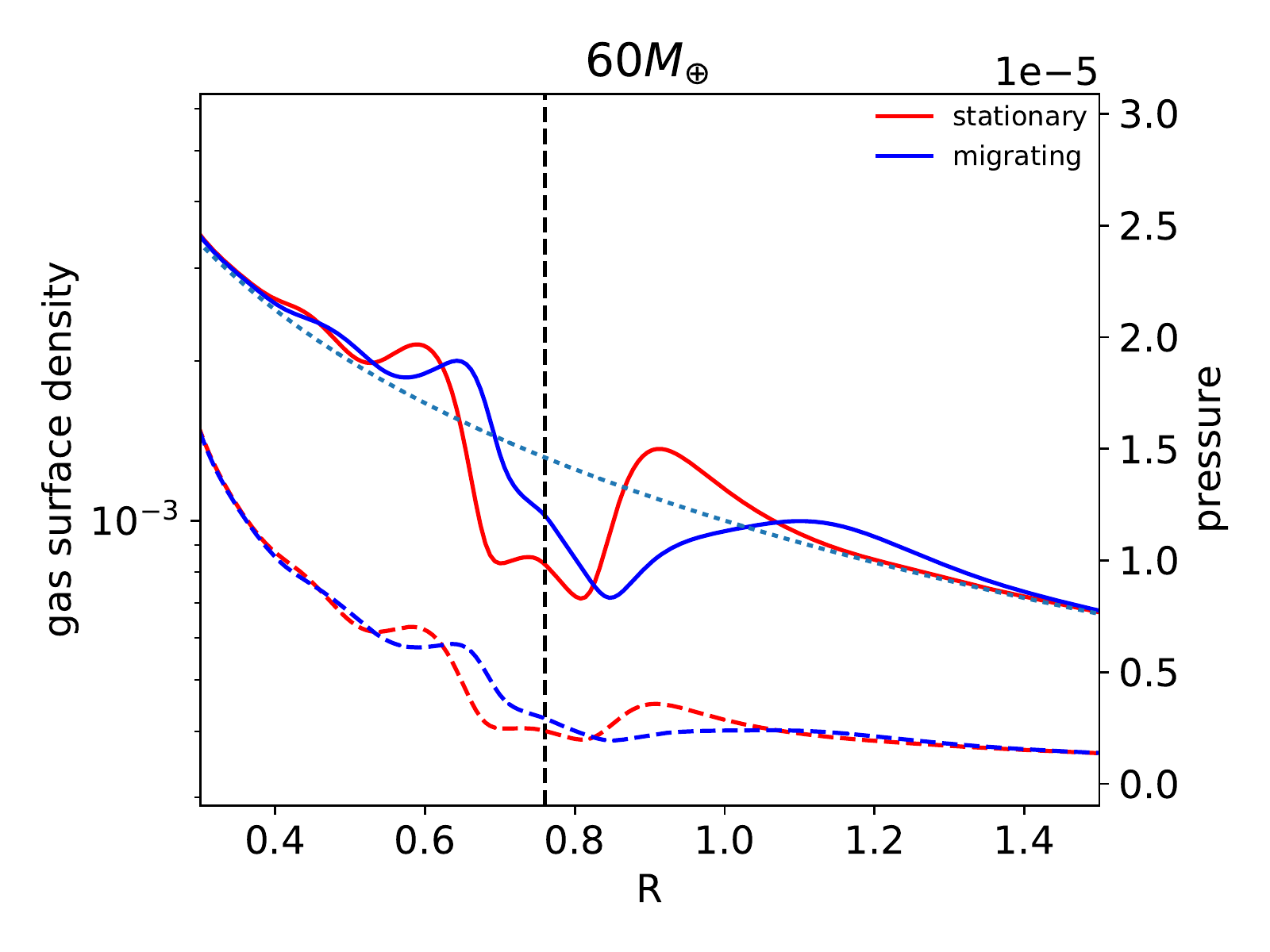}
\caption{Gas surface density (solid lines) and pressure (dashed lines) profiles for 12, 20, 30 and 60 $\MEarth$ planets at $\Rp = 0.75$.  The stationary planet is in red and the migrating planet is in blue.  The dotted line shows the initial surface density profile.  The effect of migration is firstly to reduce the effect of the planet's perturbation on the gas disc exterior to the planet, and secondly to modify the pressure perturbation interior to it for low mass planets.  The magenta line in the top right panel shows the gas profile for a planet migrating approximately three times faster than the Type I rate.}
\label{fig:1Dgas}
\end{figure*}

We model a disc that extends from $R = 0.2$ to 3 using the exact same initial surface mass density and temperature structures as \citet{Rosotti_min_detectable_Mp}.  The disc surface mass density is given by

\begin{equation}
\Sigma(R)=\Sigma_0 \frac{R_0}{R},
\label{eq:sigma}
\end{equation}
where $\Sigma_0 = 1 \times 10^{-3}$ in code units and $R_0 = 1$.  Note that the value of $\Sigma_0$ is arbitrary as the simulations are scale-free.  Nevertheless as an example, to put this into context, for a $1 \Msolar$ central star and initial planet location of 10au, this translates to $\Sigma_0 = 888 \rm g/cm^2$ at 1au.  We model a locally isothermal disc structure so that the temperature profile does not change over time, setting the aspect ratio to be

\begin{equation}
\frac{H}{R} = 0.05 \left ( \frac{R}{R_0} \right )^{0.25}.
\label{eq:h}
\end{equation}
We use the $\alpha_{\rm visc}$ viscosity prescription of \citet{SS_viscosity} with $\alpha_{\rm visc}=10^{-3}$.  Since we do not include the dust back-reaction onto the gas, the absolute normalisation of the dust density is arbitrary and the results can be rescaled to any choice of the initial dust to gas ratio. For simplicity, we use the same surface density normalisation for the dust as for the gas.

\subsubsection{Calibration of the migration rate}
\label{sec:cal_mig}
We model planets with planet-to-star mass ratios of $q = 3.6 \times 10^{-5}$, $6 \times 10^{-5}$, $9 \times 10^{-5}$ and $1.8 \times 10^{-4}$ which for a $1 \Msolar$ central star translates to planet masses of $\Mp = 12$, 20, 30 and $60 \MEarth$.  We firstly run the simulations allowing the planets to migrate naturally under the influence of the disc torques.  We measure their Type I migration timescale, $\tau_I$, by fitting the following relation that describes the decay of the planet semi-major axis as

\begin{equation}
\Rp = R_{\rm p,0} \rm e^{-t/\mbox{$\tau_I$}}
\label{eq:mig}
\end{equation}
We then repeat the simulations in a controlled way where the planet does not feel the effects of the disc but instead has its migration rate prescribed according to its measured value of $\tau_I$ in \autoref{eq:mig} (akin to the simulations performed by \citealp{Duffell_TypeII_notviscous}).  We confirm that the radial migration of the planet in the disc using this approach is similar to that of the planet migrating freely, though it is worth mentioning that \autoref{eq:mig} starts to deviate from the actual migration tracks for higher planet masses ($\gtrsim 60 \MEarth$).  While this could, in principle, affect the pressure profile for the $60 \MEarth$ planet, and hence the dust collection, the main qualitative results presented in this paper are unaffected by the use of a prescribed migration (see end of this section).  Unless specified, all simulations are performed with the planet migrating at a prescribed rate that is roughly equivalent to the Type I rate though we also run some simulations with different migration rates.  We confirm that the values of $\tau_I$ are within a factor of 2 of the values expected from linear theory, $\tau_{I,\rm lin}$, (Table~\ref{tab:tau_I}) using

\begin{equation}
\tau_{I, \rm lin} = \frac{\Rp^2 \Omega_{\rm p} \Mp}{2 \Gamma_{\rm p}},
\label{eq:tauI_lin}
\end{equation}
where $\Omega_{\rm p}$ is the Keplerian velocity evaluated at the planet's location and


\begin{dmath}
\Gamma_{\rm p} = \frac{1}{\gamma} \left [ -2.5 -1.7\beta + 0.1\alpha + 1.1 \left( \frac{3}{2} - \alpha \right ) + 7.9 \frac{\xi}{\gamma} \right ] \left ( \frac{q}{h} \right )^2 \Sigma_{\rm p} \Rp^4 \Omega_{\rm p}^2
\label{eq:torque}
\end{dmath}
\noindent is the torque exerted on the planet evaluated using equation 47 of \cite{Paardekooper_TypeI_torque}, where $\gamma$ is the ratio of specific heats, $-\alpha$ and $-\beta$ are the slopes of the disc surface mass density and temperature profiles respectively, and $\xi = \beta - (\gamma - 1)\alpha$.


Given our choice to specifically explore the effects of Type I migration, we essentially assign these simulations a scale, though note that these simulations are in fact valid for any disc mass for the migration rate prescribed in our simulations.  The main results presented in this paper involve simulations where the planet is placed in the disc at the specified mass and allowed to migrate immediately at the prescribed rate.  We also tested planets that are held on a fixed orbit before migrating as well as those allowed to freely migrate under the disc's natural torques and confirm that the qualitative result presented in this paper still holds.

\section{Results}
\label{sec:results}

\subsection{The effect of migration on the gas}
\label{sec:gas}

\begin{figure*}
\includegraphics[width=1.0\columnwidth]{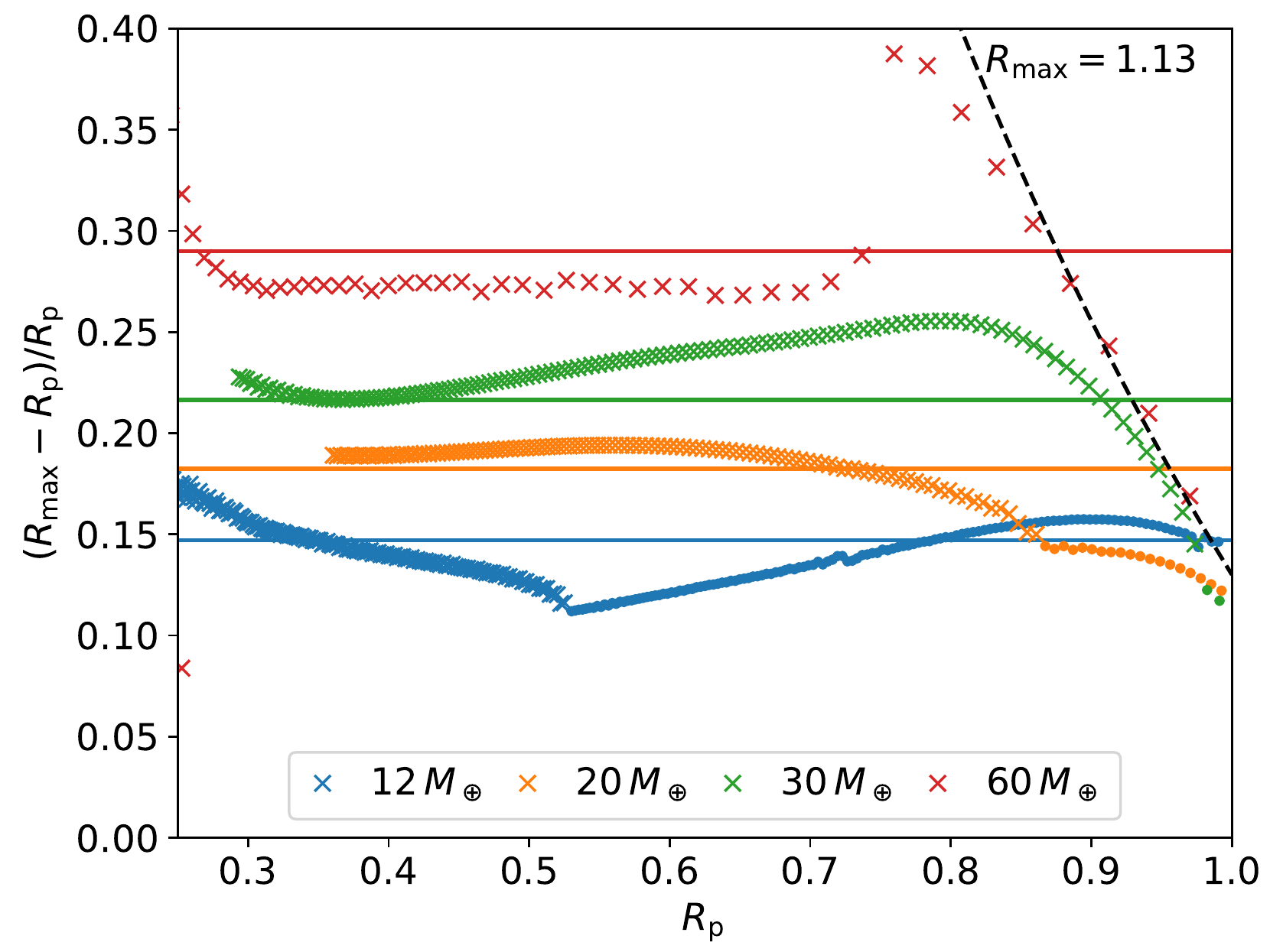}
\includegraphics[width=1.0\columnwidth]{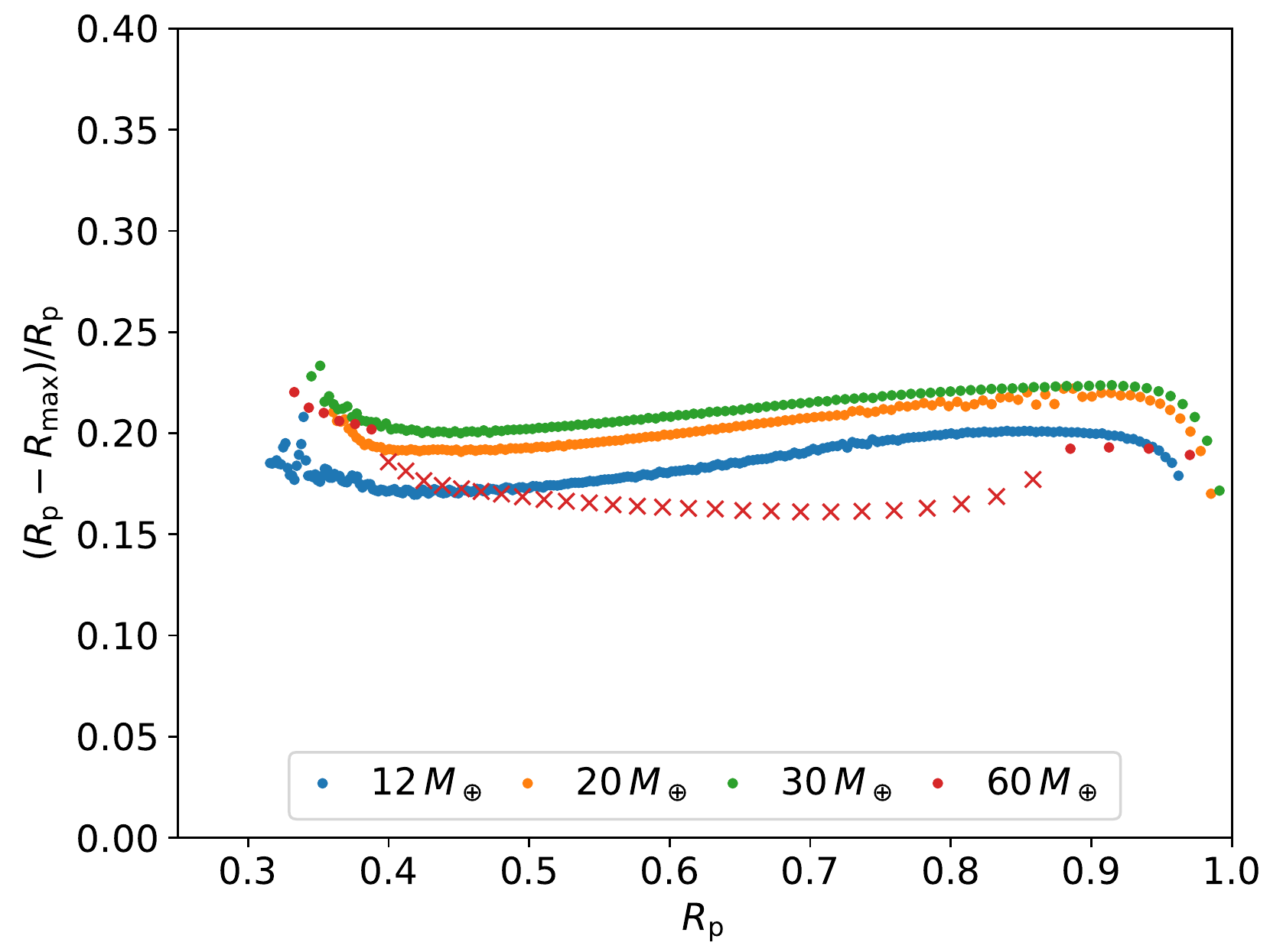}
\caption{Relative distance between the location of the planet and the pressure perturbation exterior (left) and interior (right) to the planet for various mass planets migrating from $R_{\rm p}=1$.  Crosses show times when the planet creates a full pressure maximum, while dots denote times where the magnitude of the pressure gradient is a minimum, but not zero, i.e. a point of inflection exists in the pressure profile.  The solid lines show the results determined from fits to non-migrating planets \citep{Rosotti_min_detectable_Mp} and the dashed black line shows how a fixed maximum at $R=1.13$ would appear in this plot.  For most of the simulation the pressure perturbations remain with the planet.}
\label{fig:Pmax_time}
\end{figure*}

\begin{figure*}
\includegraphics[width=1.0\columnwidth]{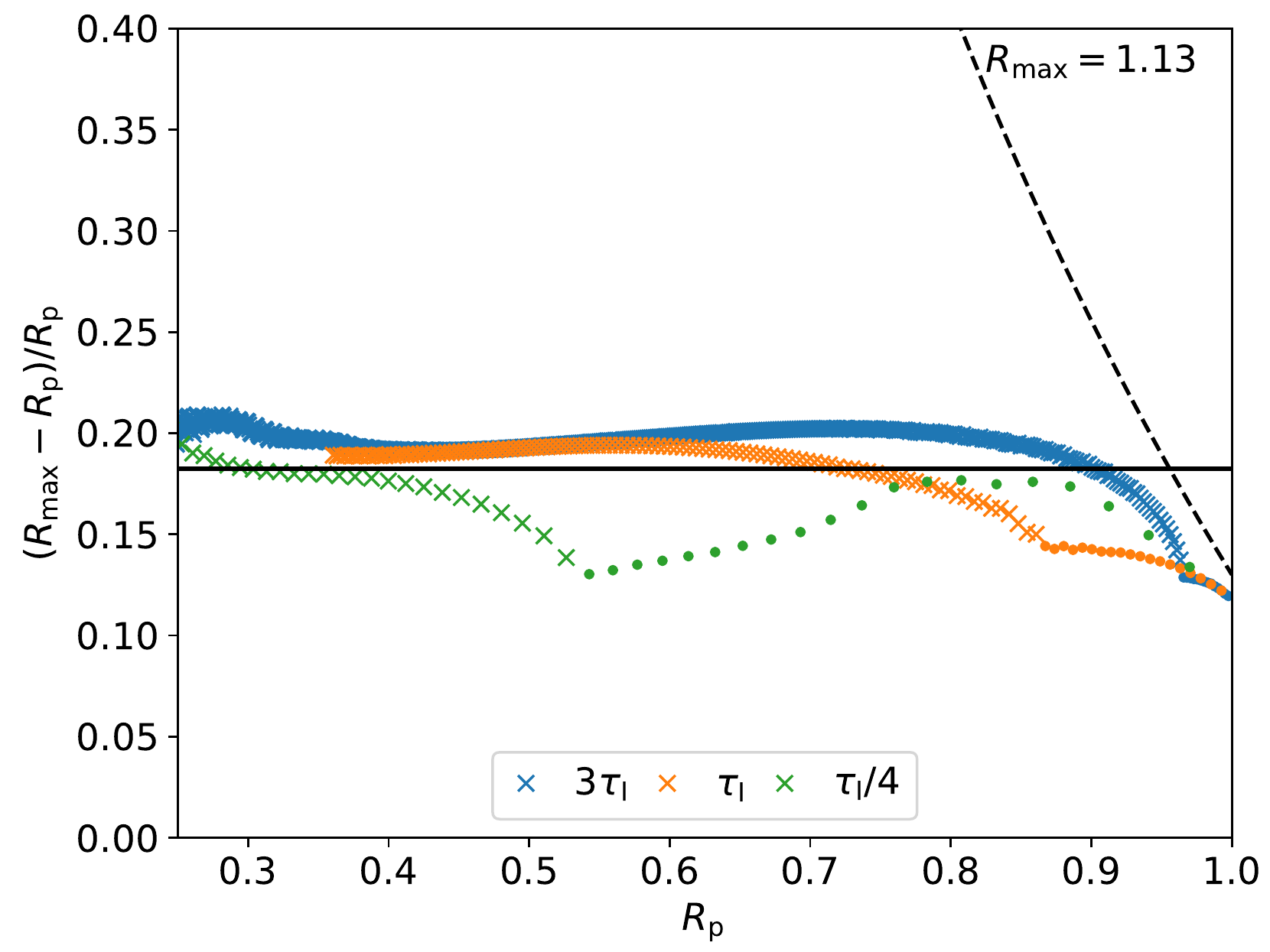}
\includegraphics[width=1.0\columnwidth]{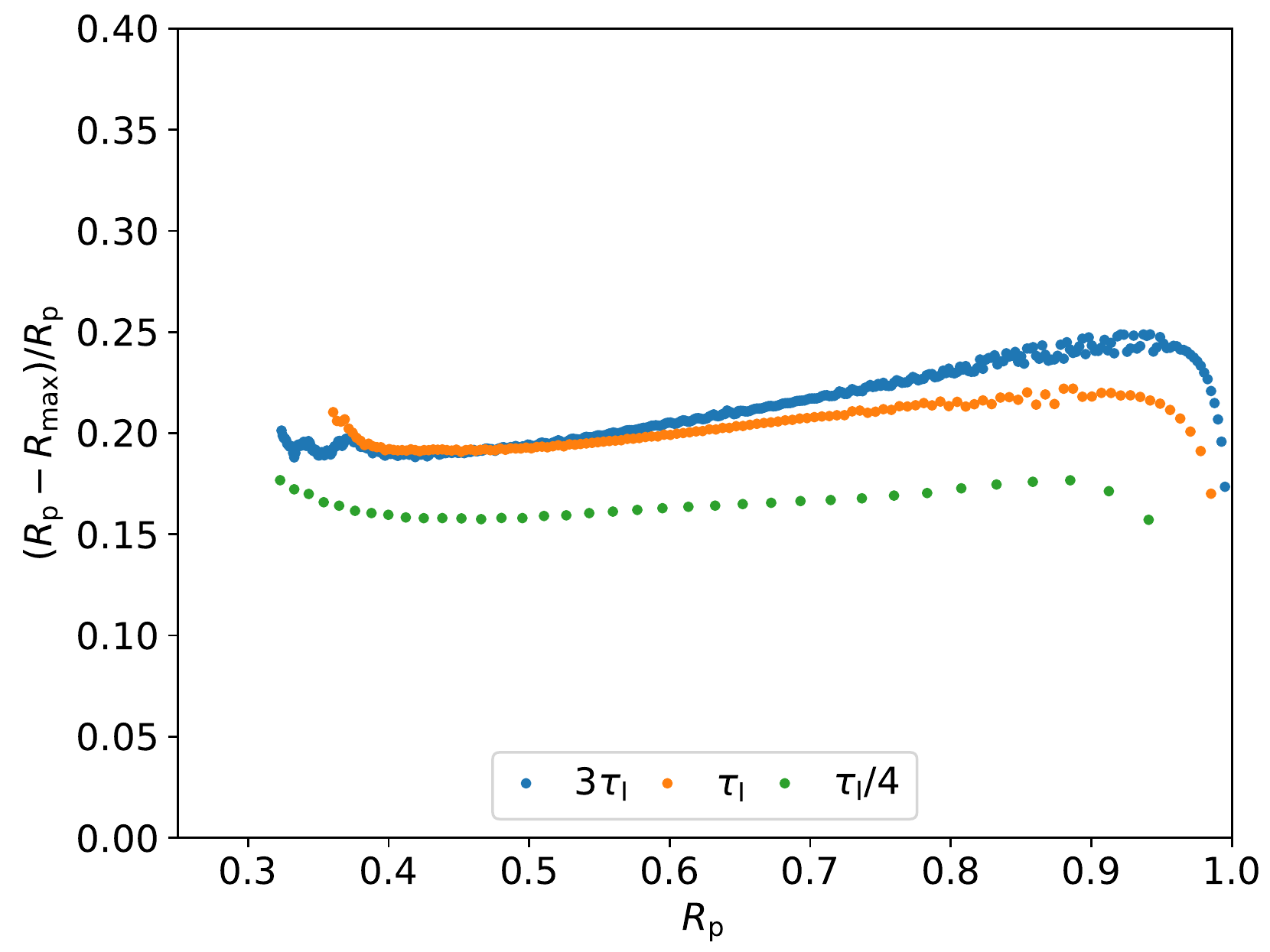}
\caption{As with \autoref{fig:Pmax_time} but for a 20 $\MEarth$ planet migrating at the Type I migration rate (green) as well as four times faster (red) and three times slower (blue).  Crosses show times when the planet creates a full pressure maximum, while dots denote times where the magnitude of the pressure gradient is a minimum, but not zero, i.e. a point of inflection exists in the pressure profile.  The pressure perturbation remains with the planet irrespective of the migration speed.}
\label{fig:Pmax_time_20}
\end{figure*}

\begin{figure}
\includegraphics[width=1.0\columnwidth]{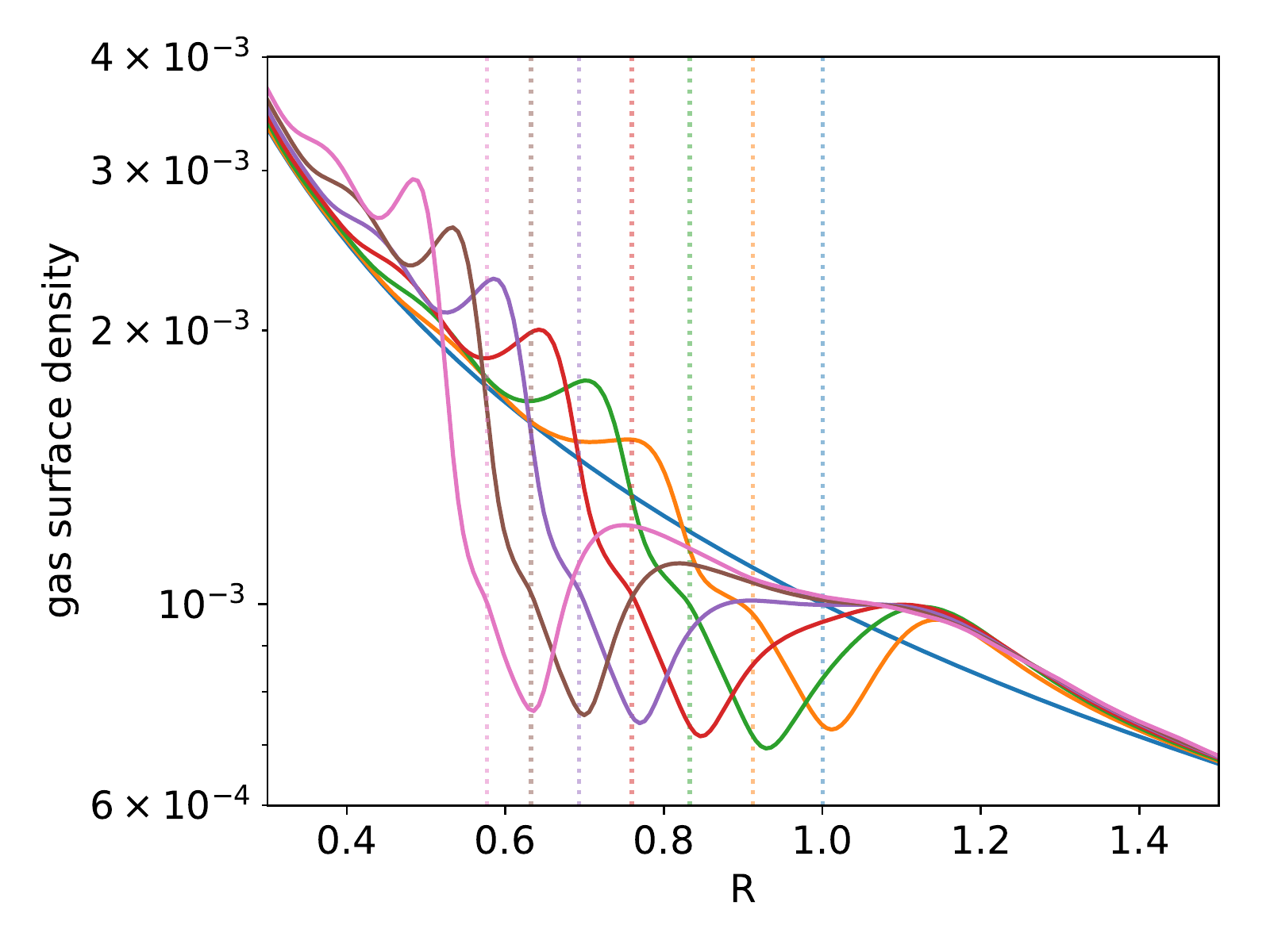}
\caption{Graph showing how the gas profile changes in the presence of a $60 \MEarth$ planet migrating at the Type I rate.  Initially a single density maximum forms but over time a second density maximum forms while the first develops into a point of inflection.  The dotted lines show the planet location.  The planet is initially at $R_{\rm p} = 1$.  The axes are in code units.}
\label{fig:1Dgas_60M_time}
\end{figure}

\begin{figure}
\includegraphics[width=1.0\columnwidth]{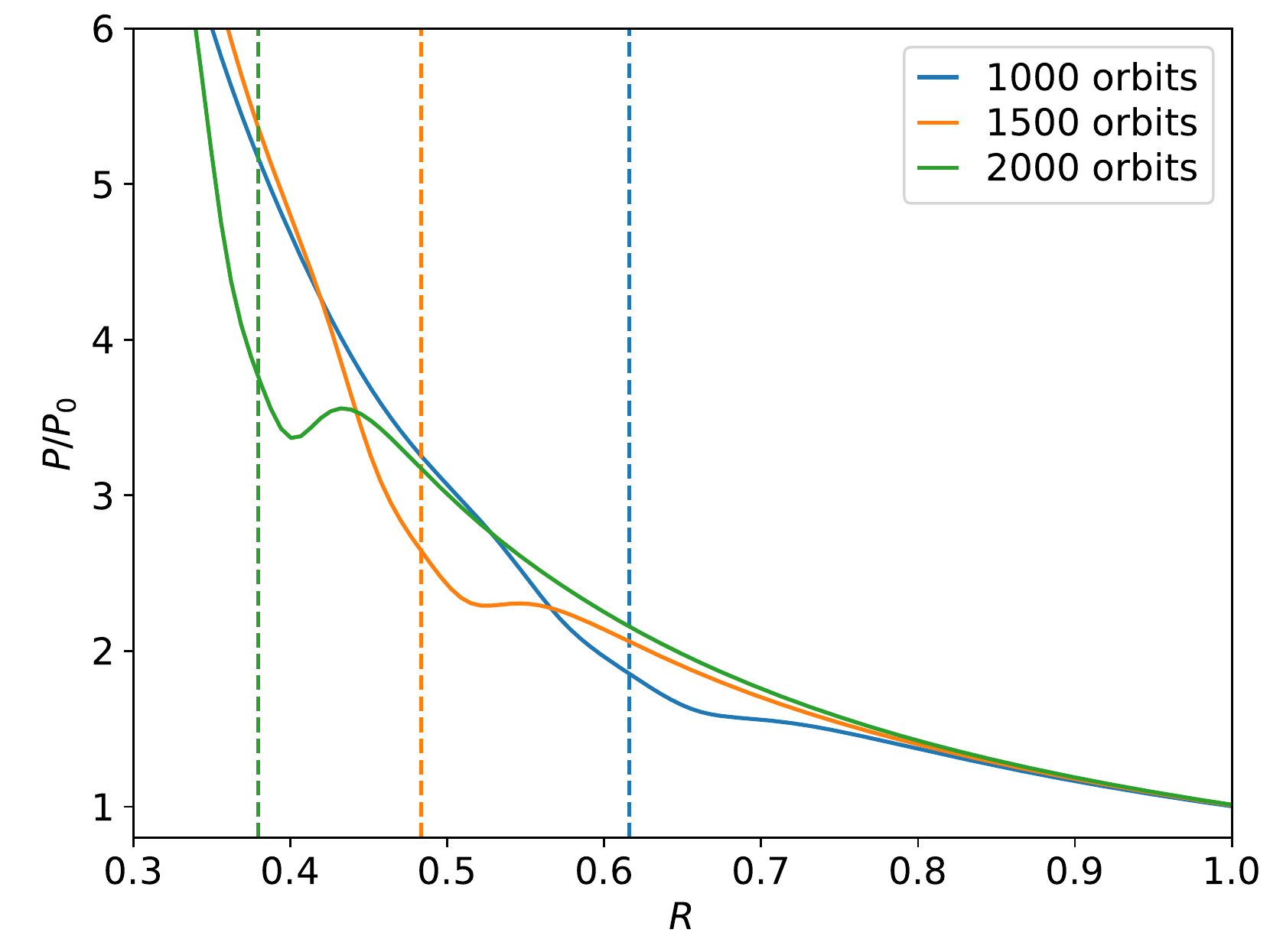}
\caption{Time evolution of the pressure profile (normalised by the unperturbed value at $R=1$) for the 12$\MEarth$ planet. As the planet migrates inwards, it transitions from creating a point of inflection (traffic jam) to a pressure maximum (dust trap) exterior to the planet due to the decrease in the disc aspect ratio (see Section~\ref{sec:gas}). The vertical dashed lines show the location of the planet.}
\label{fig:P_time}
\end{figure}

Here we consider how migration affects the pressure profile around the planet. For $\rm St < 1$ there are two ways in which the  perturbed pressure profile can give rise to rings and gaps in the dust density, as identified by \citet{Rosotti_min_detectable_Mp}.  Firstly, if the pressure perturbation is strong enough to create a local maximum in the pressure profile, then dust drifting towards the maximum will be trapped there as the sign of the radial drift velocity, given by

\begin{equation}
v_{\rm drift} = -\frac{\eta v_{\rm k}}{\rm St + St^{-1}} = \left ( \frac{H}{R} \right )^2 v_{\rm k} \pderiv{{\rm log} P}{{\rm log} R} \frac{1}{\rm St + St^{-1}},
\label{eq:vdust}
\end{equation}
changes at the pressure maximum\footnote{This equation breaks down when the deviation of the gas angular velocity from Keplerian is dominated by the planet, rather than the pressure gradient, which occurs within a few Hill radii of the planet.  This also assumes steady state, which is reasonable for $\St < 1$ \citep{Dipierro_dust_gap_criterion}}.  The Keplerian velocity is given by $v_k$.  Secondly, there can be a point of inflection where the pressure perturbation is too small to create a pressure maximum but is where the absolute pressure gradient is a minimum and thus the dust drift velocity is the slowest as the dust migrates inwards.  If the flow is in a steady state, mass conservation implies that the density will reach a maximum there. This phenomenon was termed a ``traffic jam'' by \citet{Rosotti_min_detectable_Mp}.

\autoref{fig:1Dgas} shows the gas surface density and pressure profiles for the migrating and stationary planet simulations, and shows that migration affects the pressure profile inside and outside the planet in different ways.  Firstly, the pressure maximum outside the planet is weaker.  Secondly for low mass migrating planets (12, 20 and 30 $\MEarth$) the point of inflection in the pressure profile \emph{interior} to the planet is more pronounced than in the stationary planet simulations (this is most obvious in the case of the 20 and $30 \MEarth$ planets).

\autoref{fig:Pmax_time} shows the position of the pressure maximum or point of inflection in the outer and inner disc with respect to the planet's position (and normalised by the planet's position, since the planet's semi-major axis also influences the location).  A horizontal line on this graph would show that the pressure trap moves with the planet.  With the exception of the 30 and 60 $\MEarth$ cases at the start of the simulations, the pressure traps are located within 25 percent of the distance for stationary planets \citep{Rosotti_min_detectable_Mp}, and mostly are quite close to the stationary planet estimates\footnote{At small radii the results are affected by the resolution and inner boundary.}.  Given that \cite{Rosotti_min_detectable_Mp} show that the gap width (and hence location of the pressure perturbation) scales as $M_{\rm p}^{1/3}$, this suggests that the planet mass estimates based on stationary planets are not likely to be wrong by more than a factor of two.  Note that while \cite{Rosotti_min_detectable_Mp} determined their planet mass estimates based on the locations of pressure maxima with respect to the planet, \autoref{fig:Pmax_time} shows that the mass estimates determined from a dust maximum associated with a point of inflection would be similar.  However, there is also a trend with radius, which points to some scale height dependence in this regime.  In addition we show in \autoref{fig:Pmax_time_20} that the distance of the pressure perturbation is not as sensitive to the migration speed.

In addition to the main pressure maximum that follows the planet, \autoref{fig:Pmax_time} also shows hints of a secondary pressure maximum that initially forms outside the location of the planet at $R \approx 1.13$, which can be seen by comparing the location of the pressure maximum to the $R_{\rm max} \approx 1.13$ line for the $30\,\MEarth$ and $60\,\MEarth$ planets.  Note that since we allow the planet to begin migrating immediately, the feature is not simply a residual gap or pressure maximum opened before the planet starts to migrate.  This extra pressure maximum can be seen clearly in \autoref{fig:1Dgas_60M_time}, which forms outside the initial location of the planet, slowly disappearing over time by evolving into a point of inflection.  We explain the origin of this feature in Appendix~\ref{sec:GasAndToy}.

Despite the weaker pressure maximum formed in the outer disc with migrating planets, we find that the planet mass required to produce a maximum in the pressure profile does not change dramatically for typical disc masses. \citet{Rosotti_min_detectable_Mp} found the minimum mass for creating a pressure maximum in exactly the same disc setup as ours to be between $12 \MEarth$ and 20$\,\MEarth$ at $H/R=0.05$ for non-migrating planets. Similarly, \citet{Lambrechts_halt_pebbleaccretion} reported a critical mass of $20 (H/(0.05R))^3 \,\MEarth$, typically called the `pebble isolation mass' because it is the minimum mass that forms a pressure maximum exterior to it and prevents the flux of pebbles into the inner disc.  \cite{Bitsch_PIM} and \cite{Ataiee_PIM} produce refined criteria for the pebble isolation mass considering the disc aspect ratio and turbulence.  \cite{Bitsch_PIM} also include a dependence on the disc's initial pressure profile.  Our results are consistent with the pebble isolation mass from \citet{Lambrechts_halt_pebbleaccretion} and the gas-only simulations (i.e. equivalent to what we present in this section) of \cite{Bitsch_PIM} and \cite{Ataiee_PIM}.  \autoref{fig:1Dgas} shows that the 20$\MEarth$ planet still forms an exterior pressure maximum when it migrates at the Type I rate.  However, we note that varying the migration rate can change this -- as shown by the red line in the left panel of \autoref{fig:Pmax_time_20} which shows that a migration rate a few times faster than our standard model results in a point of inflection instead of a maximum.

The flared structure used in our simulations means that the pebble isolation mass varies with radius, and all of the planets eventually create a pressure maximum in the disc once they reach small enough radii. We demonstrate this for the $12\,\MEarth$ planet in \autoref{fig:P_time} and \autoref{fig:Pmax_time} (left panel), where the transition occurs when the planet is at about $R=0.55$, where $H/R = 0.04$.   We note that while this is consistent with the pebble isolation mass from \citet{Lambrechts_halt_pebbleaccretion}, \cite{Bitsch_PIM} and \cite{Ataiee_PIM}, this cannot hold for migrating planets in general.  As an example in the $20 \MEarth$ case the radius at which the external pressure profile transitions from a point of inflection to a pressure maximum is smaller for larger migration rates (see \autoref{fig:Pmax_time_20} left panel).  This suggests that although the effect on the pebble isolation mass is rather minor in the cases studied here, this may not necessarily be the case and so an additional criterion factoring in the migration rate needs to be included into the pebble isolation mass, somewhat analogous to the gap-opening timescale criterion required for migrating planets \citep{Lin_Papaloizou_TypeII,Malik_Meru_a}, and moreso for very fast migrating planets (e.g. in the Type III regime).  Furthermore, we note that although more massive planets migrate more quickly in the Type I regime, they still open deeper gaps than less massive planets. This is a consequence of torques exerted on the disc by the planet scaling as $q^2$, while the migration speed scales with $q$.  Thus more massive planet open up deeper gaps, even when their faster migration rate is taken into account.

\subsection{The effect of migration on the dust}
\label{sec:dust}

\subsubsection{Overview of dust evolution in the presence of migrating planets}

\begin{figure}
\vspace{-0.5cm}
\includegraphics[width=1.0\columnwidth]{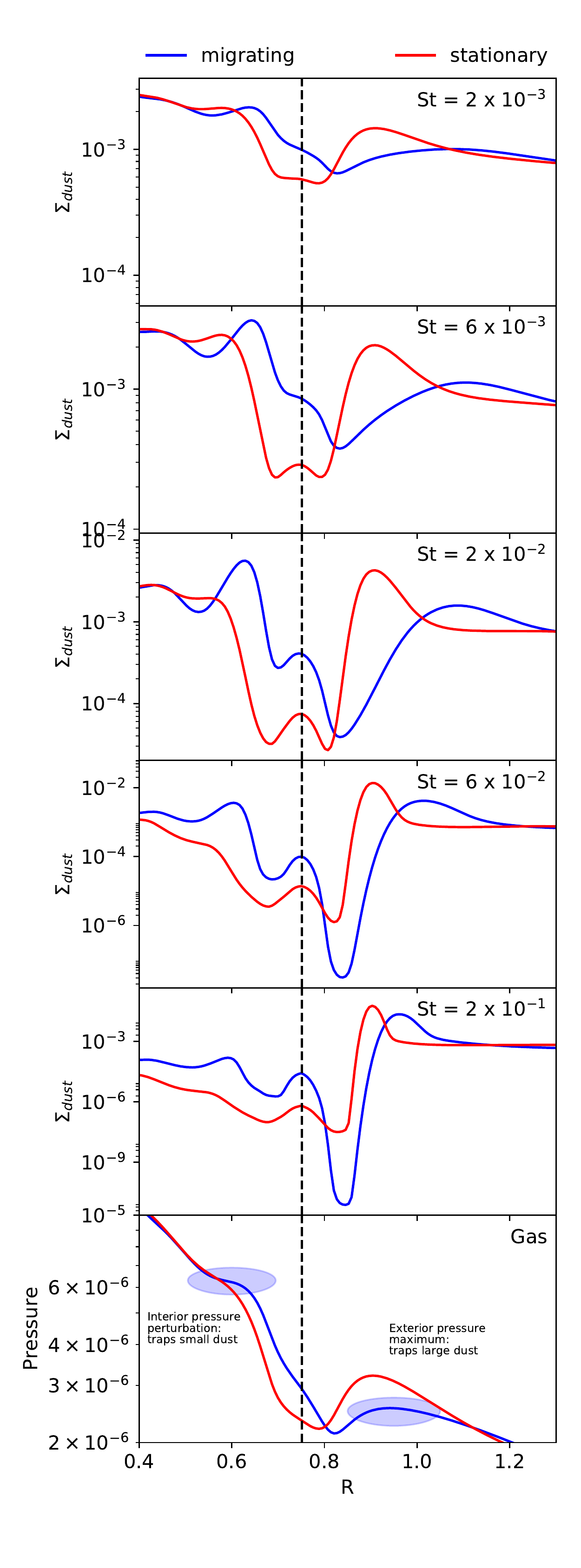}
\vspace{-1.5cm}
\caption{Dust density profile for the migrating 30 $\MEarth$ planet simulation (blue) compared to the stationary planet simulation (red).  In both cases the planet is at $R_{\rm p} = 0.75$.  The top five panels show the dust profiles for various different Stokes numbers while the lower panel shows the pressure profiles.  The migrating planet simulation shows an additional pressure trap interior to the planet in the form of a point of inflection (at $R \approx 0.6$).  The exterior pressure trap is weaker in the migrating planet simulation which results in less dust being trapped.  The snapshots of each simulation are at identical times ($t = 320$ orbits at the planet's initial location).} 
\label{fig:30MEarth_suppress}
\end{figure}

\begin{figure*}
\begin{overpic}[width=1.0\columnwidth]{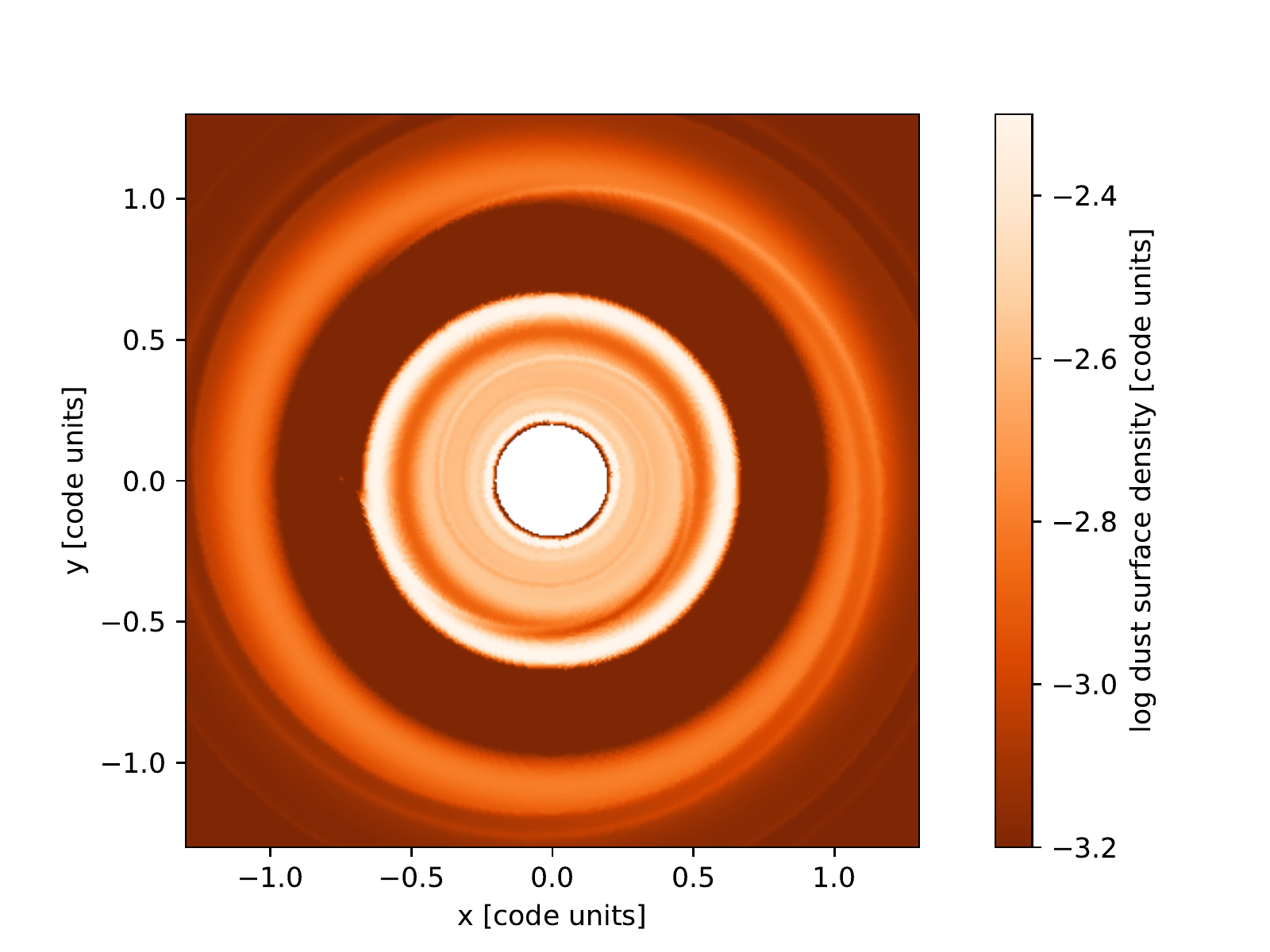}
  \linethickness{2pt}
  \put(20.75,19){\color{white}\vector(1,3){6}}
  \put(16,15){\color{white}\textsf{planet}}
  \put(59,19){\color{white}\vector(-1,2){4}}
  \put(50,15){\color{white}\textsf{inner dust ring}}
\end{overpic}
\begin{overpic}[width=1.0\columnwidth]{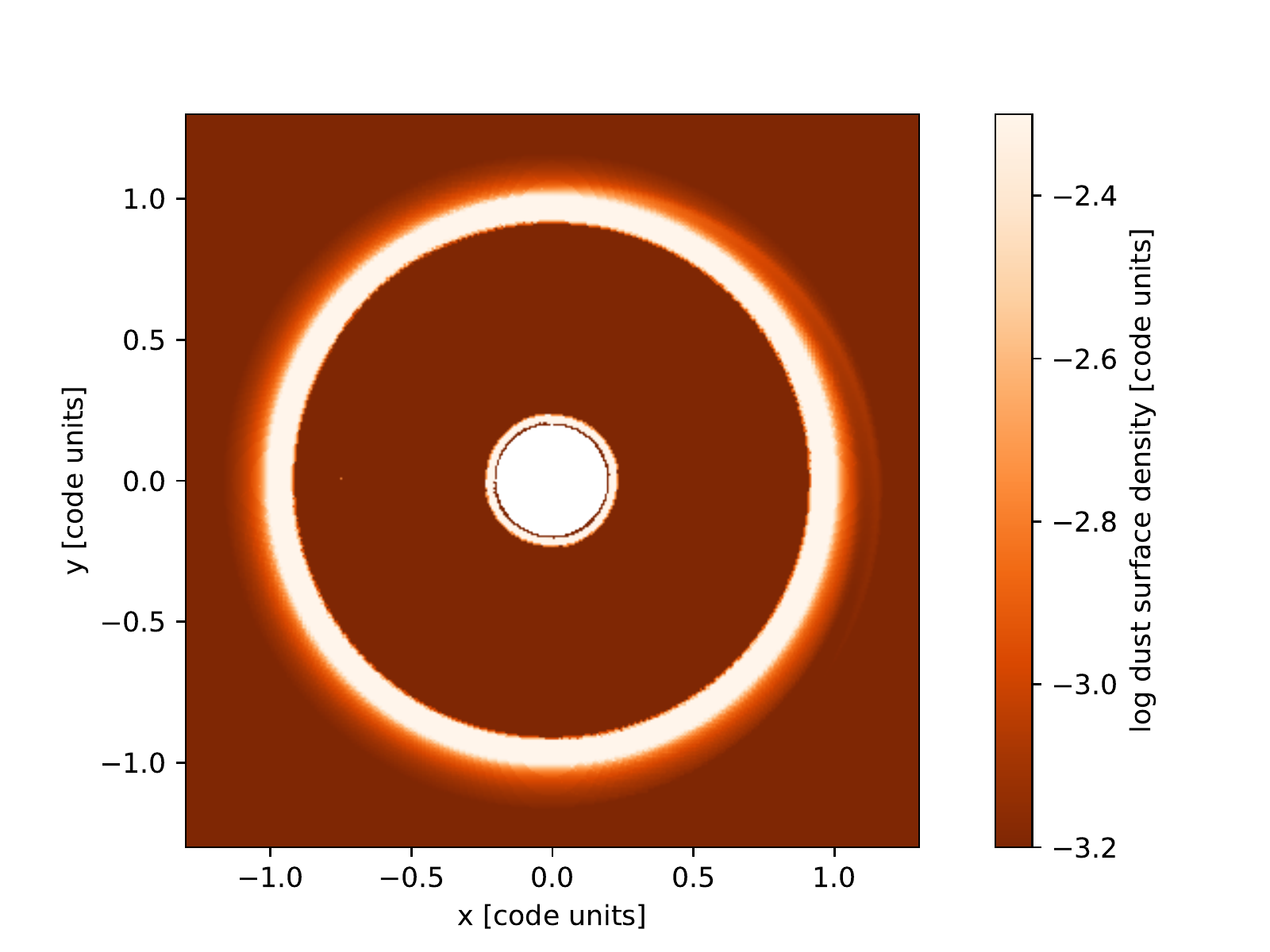}
  \linethickness{2pt}
  \put(20.75,19){\color{white}\vector(1,3){6}}
  \put(16,15){\color{white}\textsf{planet}}
  \put(64,16){\color{white}\vector(-1,2){3}}
  \put(50,12){\color{white}\textsf{outer dust ring}}
\end{overpic}
\caption{Dust density rendered simulation image of the disc with a $30 \MEarth$ migrating planet at $\Rp = 0.75$ for dust with Stokes numbers of $0.02$ (left) and $0.2$ (right).  The small dust forms a ring interior to the planet while the large dust forms a ring exterior to it.}
\label{fig:2D_disc}
\end{figure*}

\begin{figure}
\includegraphics[width=1.0\columnwidth]{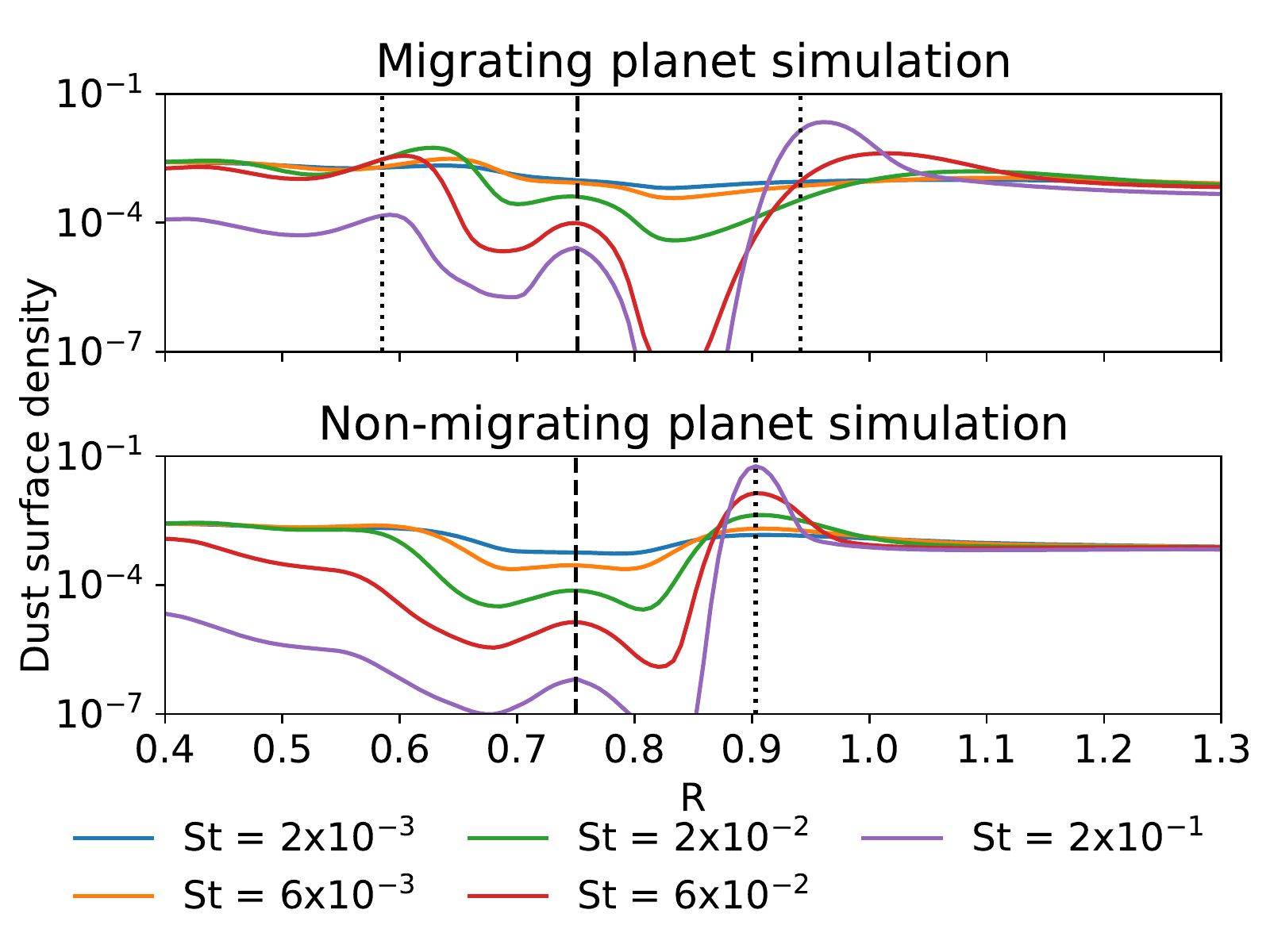}
\caption{Dust density profile for the migrating (top panel) and stationary (bottom panel) 30 $\MEarth$ planet simulations for particles of various Stokes numbers.  The dashed line shows the planet location while the dotted lines show the location of the pressure perturbations.  For stationary planets the peak in the dust occurs close to the same location for all sizes which is close to the location of the pressure maximum.  However, for migrating planets dust maxima interior and exterior to the planet do not necessarily line up exactly for different grain sizes.} 
\label{fig:30MEarth_dustpeak}
\end{figure}

\begin{figure}
\vspace{-1.75cm}
\includegraphics[width=1.0\columnwidth]{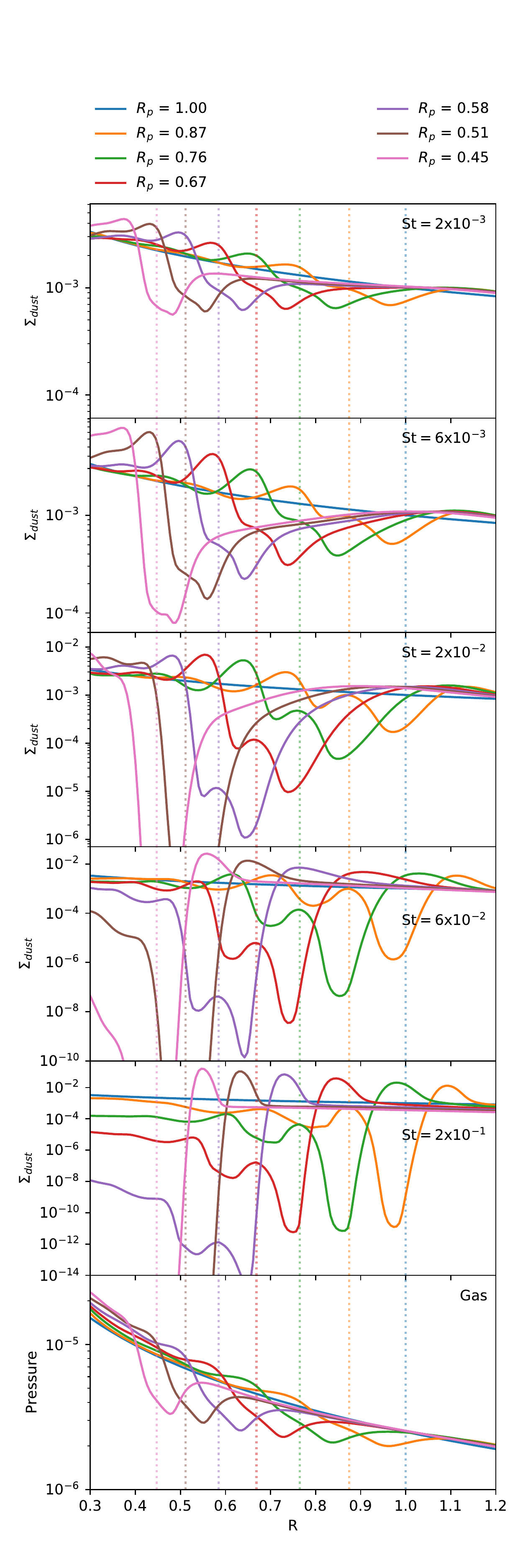}
\vspace{-1.0cm}
\caption{Time evolution of the dust density profile for the 30 $\MEarth$ migrating planet simulation for particles of various Stokes numbers (top five panels) as well as the time evolution of the pressure profile (bottom panel).  The dotted lines show the planet locations.  For smaller dust sizes the dust ring is interior to the planet while at larger sizes the dust ring is exterior to it.} 
\label{fig:30MEarth_timeevol}
\end{figure}

\autoref{fig:30MEarth_suppress} shows the surface density of the dust in the presence of a 30 $\MEarth$ planet, comparing both the migrating and stationary planet cases for each dust species considered, as well as the pressure profile.  The behaviour for the 12, 20 and 60 $\MEarth$ simulations are similar to the 30 $\MEarth$ simulation and where relevant we include additional figures in Appendix~\ref{sec:appendix}.  The bottom panel of \autoref{fig:30MEarth_suppress} and Section~\ref{sec:gas} show pressure perturbations both interior and exterior to the planet and thus dust can potentially collect in both locations in the migrating planet simulations.  This is confirmed by inspecting the dust surface density plots.  The large dust sizes always show a peak exterior to the planet.  For the smallest dust sizes there is barely a peak in the dust density exterior to the migrating planet whereas there is always a peak for the stationary planet case.  However, the dust peak interior to the planet is more prominent for the migrating planet simulations.  The presence of a ring interior to the planet in small sizes and exterior to it in large sizes is further highlighted in \autoref{fig:2D_disc}, which shows the two-dimensional dust surface density distributions for different dust sizes.

Furthermore \autoref{fig:30MEarth_dustpeak} shows that in the migrating planet simulations, different dust species have their dust density maxima at slightly different locations whereas for stationary planet simulations the dust maxima are all in a similar location and mostly coincident with the location of the pressure maximum.

The behaviour of the dust in the presence of a migrating planet is more obvious when looking at \autoref{fig:30MEarth_timeevol} which shows the time evolution of each dust size.  Firstly it is clear that for low Stokes numbers ($\St << 1$) the dust ring is interior to the planet while for larger Stokes numbers ($\St \sim 0.1$) it is exterior.  Note that an interior dust ring is not seen in the equivalent stationary planet simulations.  As the particle size increases (from $\St = 2 \times 10^{-3}$ to $\St = 2 \times 10^{-2}$) the amount of dust trapped in the inner ring increases.  As the particle size increases further the amount of dust trapped in the inner ring decreases substantially while the dust trapping in the exterior ring increases with particle size.    Comparing these graphs to the equivalent for the 12, 20 and 60 $\MEarth$ planets in Appendix~\ref{sec:appendix} the transition from a dominant inner dust ring to a dominant outer dust ring happens at smaller sizes for low mass planets.  This is because the Type I migration rate is slower for lower mass planets and so the behaviour is closer to the non-migrating case (i.e. only an exterior ring) for a wide range of Stokes numbers.  We also note that though these simulations are for planets that migrate at a prescribed rate, this key result is also seen in simulations where the planet migrates freely under the influence of the disc's torques.

\subsubsection{Explanation for the switch between inner and outer ring with grain size}
\label{sec:vdust_vpl}

\begin{figure}
\vspace{-1.2cm}
\includegraphics[width=0.985\columnwidth]{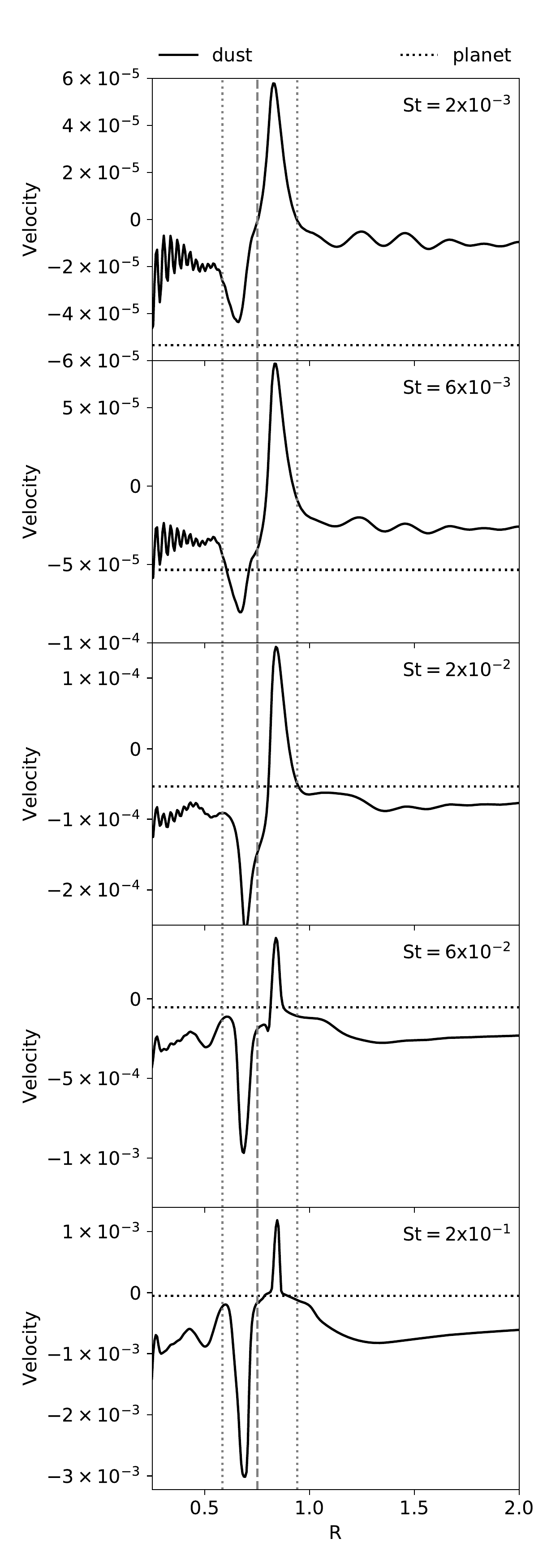}
\vspace{-1.0cm}
\caption{Dust velocity (solid line) against disc radius compared to the instantaneous planet velocity (horizontal dotted line) when the planet is at $\Rp = 0.75$ for various Stokes numbers.  The relative values of the planet velocity compared to the dust velocity (both interior and exterior to the planet) determine whether the dust ring is likely to be prominent interior or exterior to the planet.  The locations of the planet and the interior and exterior pressure perturbations are marked with vertical dashed and dotted grey lines, respectively.  Note that the oscillations in the inner disc for small Stokes numbers are transient and do not affect the key result.} 
\label{fig:vdust_vp}
\end{figure}

We can understand the switch in location of the dust density maximum by considering the relative speeds of the dust and pressure perturbations (both interior and exterior to the planet).  Since the interior and exterior pressure perturbations are moving with velocities close to that of the planet (\autoref{fig:Pmax_time}), this boils down to the relative values of the planet and dust velocities, given by $\vpl$ and $\vdust$ respectively.  \autoref{fig:vdust_vp} shows the dust velocity (determined from the simulations) for each size compared to the planet velocity when the planet is at $\Rp = 0.75$, which gives rise to the dust density profile shown in \autoref{fig:30MEarth_suppress} (blue line) and the top panel of \autoref{fig:30MEarth_dustpeak}.

The behaviour of the dust in the outer disc is different depending on whether $|v_{\rm dust}| >> |v_{\rm pl}|$, $|v_{\rm dust}| \sim |v_{\rm pl}|$ or $|v_{\rm dust}| << |v_{\rm pl}|$.  In the outer disc if $|v_{\rm dust}| >> |v_{\rm pl}|$ the dust can catch up with the moving pressure maximum or point of inflection, resulting in a dust density enhancement that we see in the stationary planet simulations (\autoref{fig:30MEarth_suppress}).  However for $|v_{\rm dust}| \lesssim |v_{\rm pl}|$ the dust cannot keep up with the pressure maximum or point of inflection so an external dust enhancement does not form.  Since the radial drift velocity is larger for particles with larger Stokes numbers, these particles have $|v_{\rm dust}| >> |v_{\rm pl}|$ and can therefore catch up with the moving pressure perturbation more easily, resulting in a higher density in the dust trap (this is the case for $\St = 6 \times 10^{-2}$ and 0.2 in the presence of a $30 \MEarth$ migrating planet).

In the inner disc if $|v_{\rm dust}| << |v_{\rm pl}|$ everywhere the dust cannot keep up with the moving pressure maximum or point of inflection.  The dust simply passes by the planet.  This occurs for our smallest dust size which has too small a radial drift velocity to keep up with the moving pressure profile.  The dust surface density of these particles closely resembles that of the gas since these dust sizes are well-coupled to it.  This is expected for these sizes since $\St \sim$ $\alpha_{\rm visc}$, i.e. the viscous flow and drift are comparable \green{\citep{Jacquet_St_alpha}}.  As the particle size is increased the dust in a small region interior to the planet travels faster than the planet due to the steep pressure gradient.  This clears a gap near the planet, pushing dust away from it.  At the same time the fast moving dust moves inwards towards the point of inflection while the slow moving dust interior to the point of inflection ends up close to the point of inflection as the planet moves inwards.  As a result the planet acts as a snowplough, collecting the dust close to the point of inflection.  This occurs for dust with $\St = 6 \times 10^{-3}$ in the presence of a $30 \MEarth$ migrating planet.  As the particle size is increased further all the dust interior to the planet travels faster than the planet.  Therefore the dust moves inwards past the point of inflection, creating a weak density maximum in the form of a traffic jam before continuing towards the star (which occurs for $\St = 2 \times 10^{-2}$, $6 \times 10^{-2}$ and 0.2).

These behaviours in the outer and inner disc occur simultaneously and it is the net effect of both of these that determines the ring location.

In summary, a pronounced outer ring is seen for higher Stokes numbers where dust is migrating fast enough to pile up in the exterior pressure maximum, as in the non-migrating case.  A pronounced inner ring is seen for low to moderate Stokes numbers when the migrating planet is driving a region of more rapidly inflowing dust into slow moving interior dust (the snowplough effect) or the pressure perturbation is such that a traffic jam occurs in the inner disc.  This is aided by the fact that these slow-moving dust particles cannot form a pronounced ring in the outer disc.  There is a narrow range of Stokes numbers for which a pronounced ring is seen both interior and exterior to the planet. This corresponds to the case where $\vdust$ in the outer disc exceeds $\vpl$ by a factor of a few and where the minimum velocity of the dust interior to the planet is $\sim \vpl$. For the $30 \MEarth$ planet this corresponds to $\rm St = 6 \times 10^{-2}$ when the planet is at $\Rp = 0.75$ but the critical Stokes number falls somewhat as the planet moves in.  For higher (lower) mass planets the critical Stokes number is higher (lower) at a given planet location.

We note the presence of small oscillations in the dust velocity at small radii for small dust sizes.  These oscillations are transient and only appear over a limited time interval in our simulations.  To investigate this further we perform a resolution test which shows that the oscillations completely disappear, and more importantly, that the key result presented in this paper remains unchanged (see Appendix~\ref{sec:appendixC} for the resolution test).

Having considered the dust dynamics at an instantaneous snapshot in time when the planet is at $\Rp = 0.75$, we can use \autoref{fig:30MEarth_timeevol} to understand the evolution of the dust rings.  The behaviour we describe earlier remains broadly the same at all times: most of the Stokes numbers remain either inner or outer ring dominated throughout the simulation.  There is, however, one exception.  The figure shows that for $\St = 6 \times 10^{-2}$ there are equally prominent inner and outer rings at an earlier time (see green line), whereas at later times, the dust density becomes increasingly outer ring dominated as the planet continues migrating inwards.  This is due to the fact that the planet velocity decreases as it migrates inwards (with our prescription, it reduces linearly with radius), while the dust radial drift velocity in the outer disc remains constant.  Therefore over time the dust moves faster than the planet, and collects in the pressure maximum exterior to it to form an outer ring in preference to an inner ring.  This is consistent with the above-mentioned velocity explanation as to why there is a switch between the inner and outer ring with grain size.  This shift in the relative dominance of the planet migration and the dust drift speed over time means that there is a small shift (of order unity) in the Stokes number at which the two rings are equally prominent.

Finally, as noted earlier, a transition from a dominant inner dust ring to a dominant outer one happens at smaller sizes for low mass planets (Appendix~\ref{sec:appendix}).  This occurs because the Type I migration rate is slower for the lower mass planets and so the smaller dust in the outer disc is able to keep up with the pressure perturbation (maximum or point of inflection).

\subsubsection{Complex gap profiles}

While the main feature associated with migrating planets is the transition between an inward lying ring to an outward lying ring as the grain size increases, there are further complexities with the gap profiles.  Firstly, it is important to note that the planet location is not in the centre of the gap as it is in the stationary planet simulation, thus affecting the estimates of the planet mass according to \cite{Rosotti_min_detectable_Mp} (see Section~\ref{sec:Mp_estimate}).

Secondly, for intermediate Stokes numbers ($\St \sim 0.01$ for the $30 \MEarth$ planet) the dust profile exterior to the planet shows an elongated profile where the dust density gently increases with radius and there is no dust enhancement close to the pressure maximum.  This occurs because the dust in that region is not being replenished: the dust in the outer disc migrates slower than or approximately equal to the planet's velocity, while the dust in the inner disc is prevented from passing by the planet either due to the snowplough effect or because it migrates in faster than the planet.  This results in the region behind the planet slowly becoming depleted in dust.

Finally, any complexities associated with the 'outer bump' seen in high planet mass simulations are discussed in Appendix~\ref{sec:60MEarth_gap}.

\section{Discussion}
\label{sec:disc}

\subsection{Observational signatures of planetary migration}
\label{sec:obs}

\begin{figure}
\includegraphics[width=1.0\columnwidth]{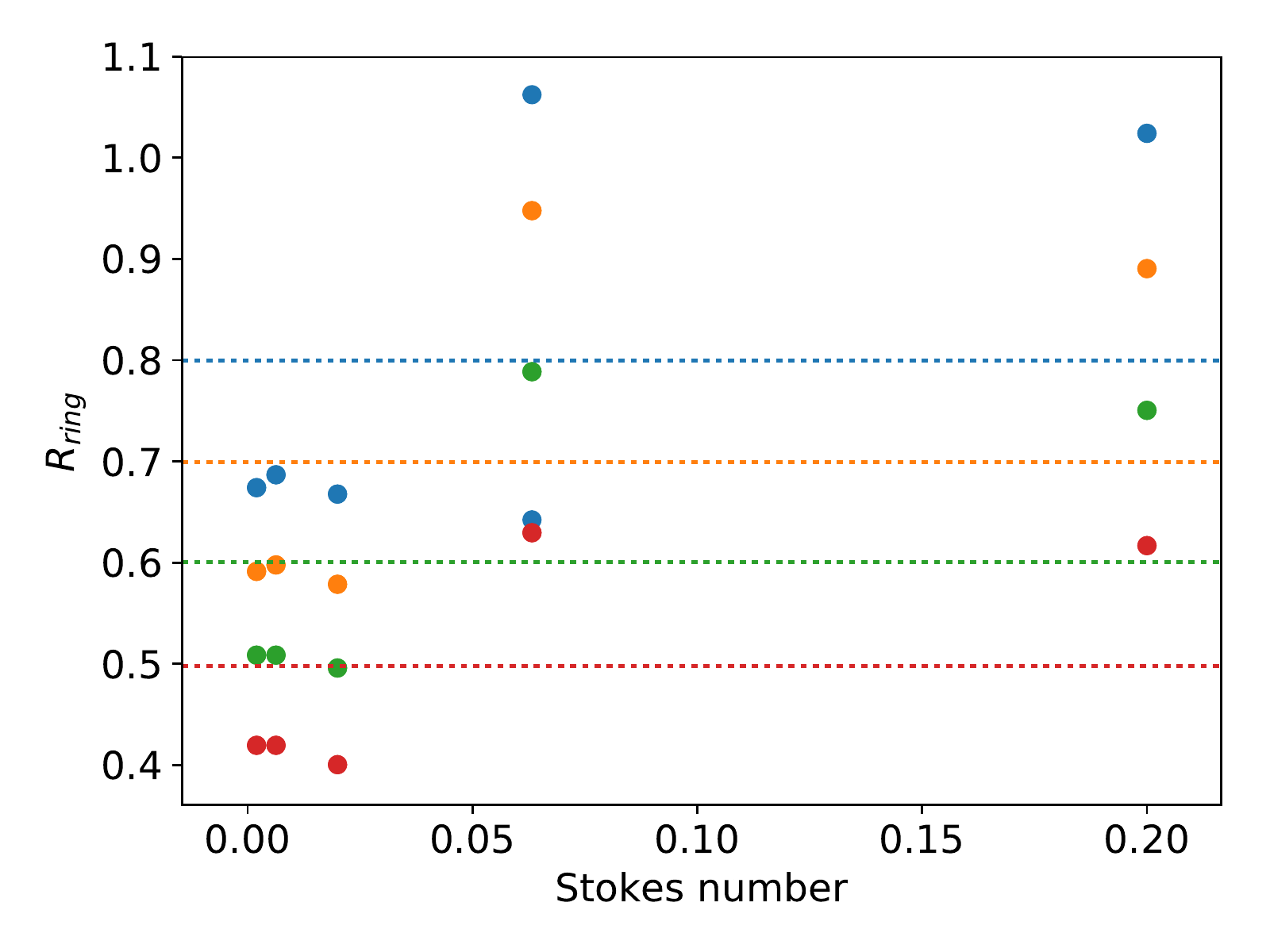}
\caption{Location of the peak dust density against Stokes number at various times in the simulation of a migrating 30 $\MEarth$ planet when the planet is at $\Rp = 0.8$ (blue), 0.7 (orange), 0.6 (green) and 0.5 (red).  The dotted lines show the location of the planet.  At small Stokes numbers the dust peak is interior to the planet and  roughly at the same location while at higher Stokes numbers the peak is exterior to the planet.  As the Stokes number (and hence dust size) increases, the location of the peak dust ring sharply moves to larger radial distances.  Note that when the planet is at $\Rp = 0.8$, the dust density interior and exterior to the planet are roughly equal for $\St = 6 \times 10^{-2}$, so two dust rings are evident (as can be seen by the orange and green lines in the fourth panel of Figure~\ref{fig:30MEarth_timeevol}).}
\label{fig:rring}
\end{figure}

Our results clearly show that the location of the ring varies with dust size with the biggest change occurring when there is a switch from a dominant inner to a dominant outer ring\footnote{This signature is distinct from the recent results of \citet{Bae_prim_sec_tertiary_arms} where planets produce multiple rings at the same locations \textit{for all} dust sizes, in low viscosity discs.}.  Consequently since the wavelength of peak emission depends on the dust size, multi-wavelength imaging of the disc midplane could in principle indicate planetary migration through a variation of ring morphology with wavelength (i.e. a shift to a more prominent outer ring at longer wavelengths).  This is illustrated in \autoref{fig:rring} which shows how the location of the dust density maximum varies with Stokes number (or equivalently, dust size) as the $30 \MEarth$ planet migrates inwards.  In particular the peak location is roughly independent of Stokes number at small dust sizes, but as the dominant peak shifts from interior to exterior to the planet, a big change in the location of the dust ring peak will be seen -- by as much as a $\approx 50$ per cent change in radial location.  As the dust size is increased further, the location of the peak once again varies only a little with dust size.  Observing such a rapid change at intermediate dust sizes may well indirectly indicate the presence of a migrating planet.  We note that in reality it is not as straightforward as this since grains of various sizes contribute to the flux at any one wavelength.  While we expect that the relative intensities of the inner and outer rings should vary with wavelength, it remains to be seen if this shift towards larger grain sizes being concentrated in the outer ring would in practice be detectable via spatially resolved observations of the spectral index in the submm regime (we are investigating this in a separate paper).  We note that based on our simulation results, the switch from an inner dominant to an outer dominant dust ring for a $30 \MEarth$ planet occurs at $\approx \rm mm$ sizes at 30au, assuming silicate grains (with density $1 \rm g cm^{-3}$) in a $0.01 \Msolar$ disc extending out to 100au, making this result particularly applicable to discs observed by ALMA (see Section~\ref{sec:threshold} for a discussion on the transition dust size).

Furthermore since an interior dust peak is never seen in stationary planet simulations, if a single gap is observed in a disc with a bright inner ring, even at a single wavelength, this is strongly indicative of a migrating planet.

\subsection{Planet mass estimates}
\label{sec:Mp_estimate}

\cite{Rosotti_min_detectable_Mp} identified several different ways of estimating the planet mass from observations.  Since they kept the planet fixed on circular orbits, it is important to consider how migration affects their estimates.  They showed that quantitative measurements of the planet mass require measuring either the gap width in scattered light images, $\Delta_{\rm gap}$ (see their Figure 16 which gives the relation $\Mp \propto (\Delta_{\rm gap}/\Rp)^{3/1.143}$), or both the position of the bright ring outside the planet in sub-mm images \emph{and} the gap width in scattered light images (see their Figure 17 which gives the relation $\Mp \propto (\Delta_{\rm ring}/\Rp)^{1/0.32}$, where $\Delta_{\rm ring}$ is the distance between the planet and the exterior ring). We notice that assessing the full impact of planet migration on these criteria would require simulated observations, which we postpone to a future paper; in what follows we simply sketch the possible implications.

Using single wavelength observations and assuming that the ring is exterior to the planet, we can try to understand the errors in the planet mass estimates if one was to use the aforementioned relations derived for stationary planets.  In Section~\ref{sec:gas} we show that the position of the pressure maximum is not significantly displaced by migration, which would lead at most to a factor of 2 difference in planet mass. It has to be noted, however, that \autoref{fig:30MEarth_dustpeak} shows that in the migrating case there is some shift between the pressure maximum location and the maximum in the dust surface density. The effect is minimal for large dust and increases for smaller sizes (see also \autoref{fig:rring}). As an example, for Stokes numbers of 0.2 and 0.06, the difference is $\approx 15$\%, i.e. comparable or slightly smaller than the shift in the pressure maximum position due to migration, which would lead to a difference in planet mass of $\approx 50$\%. Therefore, to be on the safe side one would need to observe the disc at wavelengths as long as possible in order to use the relations by \cite{Rosotti_min_detectable_Mp} to accurately determine the planet mass.

To obtain the gap width from the scattered light observations, the plots in \autoref{fig:1Dgas} show that the gap shape is affected by migration: the gap is now asymmetric, and the planet, rather than lying at the centre of the gap, is relatively close to the inner edge of the gap. For the expression obtained from Figure 17 of \cite{Rosotti_min_detectable_Mp}, the planet mass is likely to be underestimated as $\Delta_{\rm ring}$ will be underestimated while $\Rp$ will be overestimated. For the expression obtained from Figure 16 of \cite{Rosotti_min_detectable_Mp} it is unclear how this will translate quantitatively to the planet mass criterion without analysing simulated observations. Nevertheless, the fact that the planet is not in the centre of the gap is likely to have the most substantial impact on the mass estimates. Measurements of the planet mass will therefore need to fold in the fact that the planet is migrating.

An error in the planet mass calculation could also occur with the case we highlight for the $60 \MEarth$ planet, where a secondary bright ring could be produced close to the planet starting location (see Appendix~\ref{sec:appendix}).  If this maximum is incorrectly used it would lead to a serious overestimate of the planet mass, since the planet has now moved significantly far from it.

On the other hand, if the ring is interior to the planet we cannot use the relations by \cite{Rosotti_min_detectable_Mp} to estimate the planet mass.  If there is only a single ring that is interior to a gap then this is a strong indication that planet migration is happening (as stated in Section 4.1).  However if there are multiple rings and gaps in a disc it may not be so obvious whether the ring is interior or exterior to a planet.  One might then obtain not only an incorrect planet mass but also an incorrect planet location.

If a sub-mm ring is suspected to be interior to a planet, further observations at longer wavelengths (dominated by emission from grains with larger Stokes numbers) could be carried out to see if the bright ring then moves exterior to the planet location.  The properties of the newly found exterior ring could then be used with the \cite{Rosotti_min_detectable_Mp} estimates to determine a planet mass.  In this case migration can be included in the modelling to aid the accurate determination of the planet mass.

\subsection{Threshold size for interior/exterior dust ring}
\label{sec:threshold}

In Section ~\ref{sec:vdust_vpl} we show that the external density enhancement can only occur if $|v_{\rm dust}| >> |v_{\rm pl}|$ while for $|v_{\rm dust}| \lesssim |v_{\rm pl}|$ the external dust ring does not form.  Therefore an approximate criterion for the threshold size at which the switch between the dominant inner and dominant outer ring occurs is $|v_{\rm dust}| = |v_{\rm pl}|$.  Using \autoref{eq:vdust}, recognising that we are concerned with $\rm St < 1$ so that $\rm St + St^{-1} \sim St^{-1}$, and omitting order unity factors, we find that

\begin{equation}
v_{\rm dust} \approx \left ( \frac{H}{R} \right )^2 {\rm St} v_{\rm k} \pderiv{{\rm log} P}{{\rm log R}}.
\end{equation}
\autoref{eq:mig} can be differentiated to give the planet velocity

\begin{equation}
\vpl \sim \frac{\Rp}{\tau_{\rm I}},
\end{equation}
where the analytical expression for $\tau_{\rm I}$ is given by \autoref{eq:tauI_lin}.  Equating these two velocity terms gives expressions for the threshold Stokes number and dust size as

\begin{equation}
{\rm St} \sim \frac{2 | \Gamma_{\rm p} |}{\Rp \Omega_{\rm p} \Mp} \left ( \frac{H}{R} \right )^{-2} \frac{1}{v_{\rm k}} \left | \pderiv{{\rm log} P}{{\rm log} R} \right |^{-1}
\label{eq:St_threshold}
\end{equation}
and

\begin{equation}
a \sim \frac{| \Gamma_{\rm p} |}{\Rp \Omega_{\rm p} \Mp} \frac{\Sigma}{\rho_{\rm s}} \left ( \frac{H}{R} \right )^{-2} \frac{1}{v_{\rm k}} \left | \pderiv{{\rm log} P}{{\rm log} R} \right |^{-1},
\label{eq:a_threshold}
\end{equation}
respectively.  Note that since this is essentially a comparison of the planet velocity and the velocity of the dust being supplied to the pressure perturbation external to the planet, the quantities marked with a subscript 'p' are evaluated at the planet's location while all other quantities are evaluated further out in the disc.  Using equation~\ref{eq:torque} and assuming surface mass density and temperature profiles of $R^{-1}$ and $R^{-1/2}$, respectively, as modelled in this paper, these expressions become

\begin{equation}
{\rm St} \sim 0.03 \left ( \frac{\Sigma}{5 \rm g/cm^2} \right ) \left ( \frac{R}{30 \rm au} \right )^{2} \left ( \frac{\Mp}{30 \MEarth} \right ) \left ( \frac{M_{\star}}{\Msolar} \right )^{-2} \left ( \frac{H/R}{0.04} \right )^{-4}
\label{eq:St_threshold_normalised}
\end{equation}
and

\begin{align}
a \sim 0.7 \left ( \frac{\Sigma}{5 \rm g/cm^2} \right )^2 \left ( \frac{R}{30 \rm au} \right )^{2} \left ( \frac{\Mp}{30 \MEarth} \right ) & \left ( \frac{M_{\star}}{\Msolar} \right )^{-2} \left ( \frac{H/R}{0.04} \right )^{-4} \nonumber \\
& \times \left ( \frac{\rho_{\rm s}}{1 \rm g/cm^3} \right )^{-1} \rm mm
\label{eq:a_threshold_normalised}
\end{align}
\noindent where $\rho_{\rm s}$ is the internal density of the dust grains.  As mentioned in Section~\ref{sec:obs}, this critical grain size at which the ring should switch between interior and exterior to the planet (assuming silicate grains with density $1 \rm g cm^{-3}$ at a radial location of 30 AU in a $0.01 \Msolar$ disc extending out to 100au with a surface density slope of $\Sigma \propto R^{-1}$, aspect ratio given by $H/R = 0.04 (R/R_0)^{1/4}$ and a $30 \MEarth$ planet) is $a \approx \rm mm$.  \autoref{eq:a_threshold} has a quadratic dependence on the disc surface mass density, a quartic dependence on the disc aspect ratio and a linear dependence on the planet mass, this expression for the threshold dust size is highly dependent on the disc and planet properties.  We evaluate \autoref{eq:St_threshold_normalised} for our $30\MEarth$ simulation and find that the threshold Stokes number is $\approx 0.022$.  This analytically determined expression is in agreement with our simulations which show a threshold between $\St = 0.02$ and 0.06 (\autoref{fig:30MEarth_timeevol}).

\subsection{Caveats}
\emph{Initial setup:}  In our simulations we migrate the planet as soon as we place it in the disc, while often in planet-disc interaction simulations, planets are held on a fixed circular orbit at the initial location for several orbits before being allowed to migrate.  However we choose not to do this as we do not want the planet to artificially create a pressure trap as a result of this and we want to see that any dust trapping that occurs is real in the presence of a migrating planet.  Note that the feature of the calculations that is particularly sensitive to this initial set-up is the extra bump exterior to the planet.  However this only has a significant effect on the dust density distribution in the case of the $60 \MEarth$ planet.\\

\noindent \emph{Planet growth:}  We also do not grow the planet's mass over time while in reality the planet would grow slowly initially and then undergo runaway accretion.  Ideally self-consistent calculations with a planet growing at a realistic rate that starts to migrate as it grows (assuming the growth timescales are not slow compared to the migration timescale) are needed to determine the robustness of these results.  However it is important to note that the regime that we explore, i.e. without accretion onto the planet, is still relevant because it has been shown that i) the critical core mass for runaway gas accretion can be as much as $60 \MEarth$ \citep{Rafikov_60MEarth_crit_Mcore}, and ii) it is no longer clear what conditions runaway accretion occurs under \citep{Ormel_crit_Mcore1,Cimerman_core_Mcrit3,Ormel_crit_Mcore2,Lambrechts_core_Mcrit}, and iii) there are many rocky super-Earths with only a 5-10\% hydrogen envelope (in mass), suggesting that gas accretion is slow \citep{Hadden_Lithwick2017}.\\

\noindent \emph{Back-reaction:}  We do not include back-reaction of the dust on the gas in these simulations.  While this can be potentially important for dust-to-gas ratios of $\sim 1$, for our parameters back-reaction only significantly affects the gas velocity for $\St > 0.1$ and thus we do not expect it to be a major factor (see Appendix~\ref{sec:appendixB}).  We point out, though, that in cases where the dust-to-gas ratio becomes large, detailed modelling should include this effect, which may be important for interpreting observations.\\

\noindent \emph{Dust growth:}  Finally since the dust growth time-scale is comparable to or shorter than the radial drift timescale \citep{Birnstiel_growth_model}, to complete the picture of dust evolution one should consider the growth and fragmentation of dust in the simulations.\\

\noindent \emph{Migration direction:}  This study only considers the effect that inward migration has on dust rings when planets migrate at the Type I rate.  It is possible that for part of a planet's radial evolution, it may migrate in an outwards direction \citep{Lyra_Klahr_outwards_zero_torque,Ayliffe_Bate_outward_RT,Ayliffe_Bate_outward_tempgradient,Bitsch_Kley_ecc_radiative,Bitsch_Kley_outward,Bitsch_stellar_irradiation_equilibrium,Bitsch_stellar_irradiated_accretion,Bitsch_stellar_irradiation_viscosity,Paardekooper_outward_mig,Paardekooper_Mellema_outwards2} or end up in a zero torque location with no net migration with respect to the rest of the disc \citep{Lyra_Klahr_outwards_zero_torque,Bitsch_Kley_outward,Bitsch_stellar_irradiation_equilibrium,Bitsch_stellar_irradiated_accretion}.  If the planet ends up in a zero torque location, we expect that the formation and evolution of dust rings follow what we already know from stationary planet simulations: i.e. that a ring only forms exterior to the planet's location.  In the case of outwards migration we expect the ring to form exterior to the planet for all dust sizes considered, though further studies are needed to address this scenario.

\section{Conclusions}
\label{sec:conc}

We perform two dimensional hydrodynamical simulations using gas and dust to understand the formation and evolution of dust rings in the presence of low mass (12-60 $\MEarth$) migrating planets.  We find that the gas pressure profile is significantly different compared to that in the presence of stationary planets: the pressure perturbation exterior to the planet is weaker while that interior to the planet becomes more important for migrating planets.  Dust can therefore be enhanced both interior or exterior to the planet and the result is governed by the relative values of the planet and dust velocities.  For small sizes, the dust velocity in the outer disc is too small to keep up with the moving pressure maximum while in the inner disc it moves faster and can collect forming a dust density enhancement interior to the planet.  On the other hand, for large sizes, the dust velocity in the exterior disc is large enough to keep up with the pressure perturbation (maximum or point of inflection), resulting in a dust density enhancement exterior to the planet.  Consequently the location of the dust ring varies for different dust sizes, varying by as much as $\approx 50$ per cent.  There is also an intermediate size where the density enhancement is comparable interior and exterior to the planet.  The switch between a dominant inner to a dominant exterior bright ring occurs when the velocity of the dust in the outer disc is roughly equivalent to the planet velocity.  We predict that the location of the dust ring in the disc midplane observed at different wavelengths will shift outwards significantly at a particular wavelength.  Therefore in principle, it may be possible to use the location of dust rings in order to detect planetary migration, although the feasibility of this measurement is yet to be established.

\section*{Acknowledgements}
FM acknowledges support from The Leverhulme Trust, the Isaac Newton Trust and the Royal Society Dorothy Hodgkin Fellowship.  GR, RB and CC are supported by the DISCSIM project, grant agreement 341137 funded by the European Research Council under ERC-2013-ADG.  This work was undertaken on the COSMOS Shared Memory system at DAMTP, University of Cambridge operated on behalf of the STFC DiRAC HPC Facility. This equipment is funded by BIS National E-infrastructure capital grant ST/J005673/1 and STFC grants ST/H008586/1, ST/K00333X/1.  This work also used the DiRAC Data Centric system at Durham University, operated by the Institute for Computational Cosmology on behalf of the STFC DiRAC HPC Facility (www.dirac.ac.uk). This equipment was funded by a BIS National E-infrastructure capital grant ST/K00042X/1, STFC capital grant ST/K00087X/1, DiRAC Operations grant ST/K003267/1 and Durham University. DiRAC is part of the National E-Infrastructure.  This research was also supported by the Munich Institute for Astro- and Particle Physics (MIAPP) of the DFG cluster of excellence "Origin and Structure of the Universe".  This paper made use of NumPy \citep{Numpy}, SciPy \citep{scipy} and Matplotlib \citep{Matplotlib}.




\bibliographystyle{mn2e}
\bibliography{allpapers}




\appendix

\section{Toy model for the outer gas bump caused by high mass planets}
\label{sec:GasAndToy}

\begin{figure}
\includegraphics[width=1.0\columnwidth]{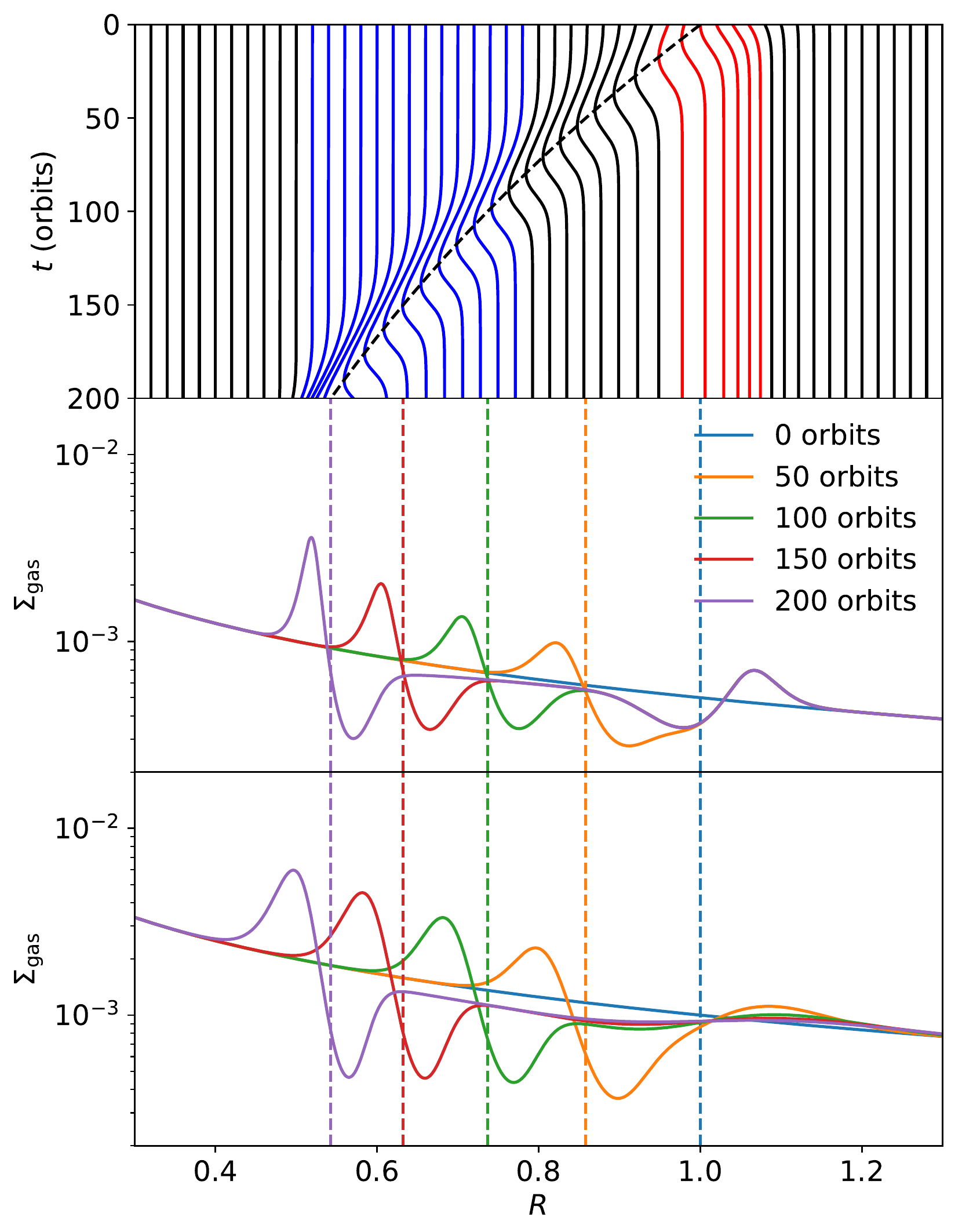}
\caption{Toy model developed to explain the outer pressure bump seen in the $60\,\MEarth$ migrating planet simulation. Upper panel: The Lagrangian trajectories of fluid elements moving under the torque (solid lines, \autoref{eq:ToyTorque}). The dashed line shows the location of the planet.  Blue, black and red lines indicate net inward, zero and outward motion, respectively.  Middle panel: the resulting surface density evolution computed from the stream line integration. Bottom panel: Surface density for the full model including viscosity with $\alpha_{\rm visc} = 10^{-3}$.  The toy model naturally explains the formation of the inner and outer surface density bump.  The latter diffuses over time due to viscosity.  In each panel the dashed lines show the location of the planet.}
\label{fig:toy}
\end{figure}

Here we investigate the origin of the second pressure maximum, which appears just outside the initial location of the highest mass planet (\autoref{fig:1Dgas_60M_time}) with a toy model designed to qualitatively explain the behaviours seen in the simulations.  We consider the influence of the torques exerted by the planet on the disc modelled in the framework of a 1D viscous evolution model.

We assume that the angular momentum evolution of the disc driven by the planet can be modelled locally, and with a torque profile that is not too different to the linear one acting on the planet.
For the torque density profile we use
\begin{equation}
\Lambda(R,R_{\rm p}) = f(\Delta) \Omega_{\rm p}^2 R_{\rm p}^2 q^2 \left(\frac{R_{\rm p}}{H_{\rm p}}\right)^4,
\label{eq:ToyTorque}
\end{equation}
where $q$ is the planet to star mass ratio, ${\Delta = (R - R_{\rm p}) / 0.9H_{\rm p}}$, and
\begin{equation}
f(\Delta) =  0.075 \left[\exp\left( - \frac{(\Delta - 0.2)^2}{2}\right) - \exp\left( - \frac{(\Delta + 0.2)^2}{2}\right)\right].
\end{equation}
Quantities with a subscript 'p' are evaluated at the location of the planet. This profile has a similar shape to the torque density exerted by the disc on the planet found by \citet{DAngelo_Lubow2010}, but the factor $0.075$ results in an amplitude that is about a factor of two smaller. Assuming that the gas remains on circular orbits and neglecting viscosity, the torque gives rise to a radial velocity of the gas,
\begin{equation}
v_r = \frac{2 \Lambda(R, R_{\rm p})}{ R \Omega}.
\end{equation}
In the toy model we take $H/R = 0.05$ and the planet mass to be $60\,\MEarth$.

The top panel of \autoref{fig:toy} shows the evolution of radius of different fluid elements under this torque, assuming the planet migrates according to \autoref{eq:mig} with a migration timescale of 327 orbits, and the middle panel shows the resulting surface density. This shows that torques from the planet naturally create both the gap near the planet and the secondary maximum outside of the planet's initial location. The upper panel demonstrates why the interior maximum and secondary maximum external to the planet form: most fluid elements that are initially inside the planet's location (highlighted in blue) experience a net inward migration.  This is because they either only move inwards, or first undergo an inward migration until the planet catches up with them then undergo an outward migration after the planet has passed.  However, those elements that start very close and interior to the planet location or just outside it only undergo a net outward migration (highlighted in red), resulting in a large net outward motion. Since those fluid elements that are initially located far outside the location of the planet hardly move at all, this inevitably leads to the formation of an additional bump in the density exterior to the planet.  This effect is not seen for lower mass planets simply because it is weaker: the torque depends on $q^2$, while the Type I migration timescale is $\propto q^{-1}$, resulting in the net motion of the gas scaling as $q$. Similarly, the gap depth in the toy model increases as the planet migrates because the migration time of the \emph{gas} scales as $R/v_r \propto \Omega_p^{-1}$, while we have taken the migration timescale of the planet to be fixed. Thus the net motion of the gas is relatively larger at smaller radii.

Finally, in the bottom two panels we show the surface density with the effects of viscosity included. To do this we solve the viscous evolution equation with the addition of the torque from the planet:
\begin{equation}
\pderiv{\Sigma}{t} = \frac{1}{R} \pderiv{}{R} \left[3 R^{1/2} \pderiv{}{R}(\nu \Sigma R^{1/2}) - 2 \Lambda(R, R_{\rm p}) \Sigma \Omega^{-1}\right],
\end{equation}
taking $\nu = \alpha_{\rm visc} c_{\rm s} H$ and $\alpha_{\rm visc} = 10^{-3}$. This shows that the addition of viscosity causes the secondary maximum to spread and weaken over time, and also reduces the depth of the primary planet induced gap.  Note that the absence of the extra structure in the gas density close to the planet in the toy model is evidence that the shape of the real torque distribution differs considerably from the toy model, likely in part due to the angular momentum deposition being non-local.

We note that 1D viscous evolution models have been explored in the context of gap opening by giant planets \citep[e.g.][]{Alexander_Armitage2009,Ercolano_Rosotti2015}, and it is known that while they qualitatively describe the process of gap opening (and the resultant Type II migration), they do not accurately reproduce the gap profiles \citep{Crida_gap,Kanagawa_etal2015,Hallam_Paardekooper2017}. These discrepancies arise for two reasons: firstly, the torque densities used are typically based upon the linear torques, which are modified in the presence of a gap due to the shifting of resonances \citep[see][]{Petrovich_Rafikov2012}. Secondly, the torque exerted on the disc by the planet is assumed to follow the same radial profile as the torque the disc exerts on the planet. This fails in detail because the primary effect of the planet is to excite density waves, which carry the angular momentum away. It is the damping of these waves that transfers angular momentum to the disc, with the damping often assumed to happen locally where the waves are excited. \citet{Rafikov2002} showed that in the inviscid case these waves can propagate for several scale heights before shocking, thus depositing their angular momentum, which explains the opening of double gaps by low mass planets in inviscid discs \citep[e.g.][]{Zhu_2D_3D_dust_gaps}. However, the action of viscosity causes the waves to damp earlier \citep{Takeuchi_etal1996,Dong_etal2011}, even for $\alpha_{\rm visc} = 10^{-4}$. The full complexity of the problem means that constructing an accurate model is beyond the scope of this paper.

Finally, we note that the formation of the outer bump occurs because more massive planets open a significant gap before they migrate.  If growth occurs slowly while the planet migrates, such a feature may not be seen.  However, it is worth noting that there are scenarios in which growth can be more rapid than migration, such as during pebble accretion \citep[e.g.][]{Bitsch_PA_evolving_discs}, or perhaps during the early phases of runaway gas accretion, before the planet has opened up a gap \citep[e.g.][]{Machida_pp_accretion}.  One of these scenarios could potentially be responsible for the formation of a secondary gas bump.

\section{Dust profiles for the 12, 20 and 60 $\MEarth$ simulations}
\label{sec:appendix}

\begin{figure}
\vspace{-1.75cm}
\includegraphics[width=1.0\columnwidth]{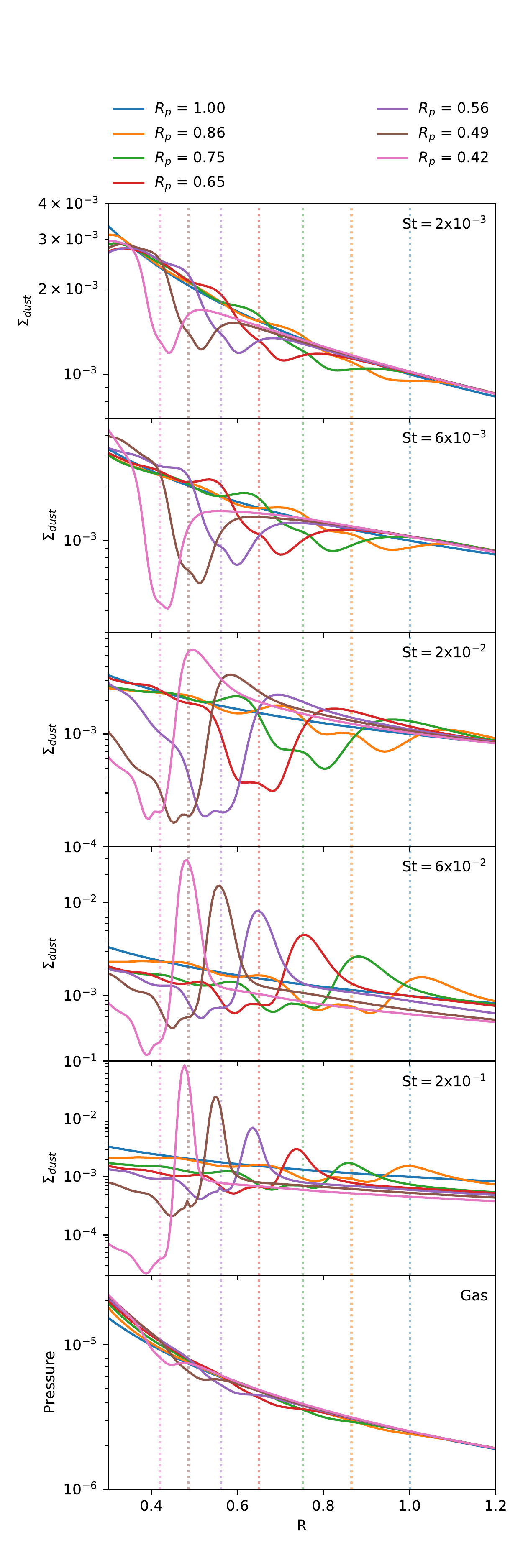}
\vspace{-1.0cm}
\caption{Time evolution of the dust density profile for the 12 $\MEarth$ migrating planet simulation for particles of various Stokes numbers (top five panels) as well as the time evolution of the pressure profile (bottom panel).  The dotted lines show the planet locations.  For smaller dust sizes the dust ring is interior to the planet while at larger sizes the dust ring is exterior to it.} 
\label{fig:12MEarth_timeevol}
\end{figure}

\begin{figure}
\vspace{-1.75cm}
\includegraphics[width=1.0\columnwidth]{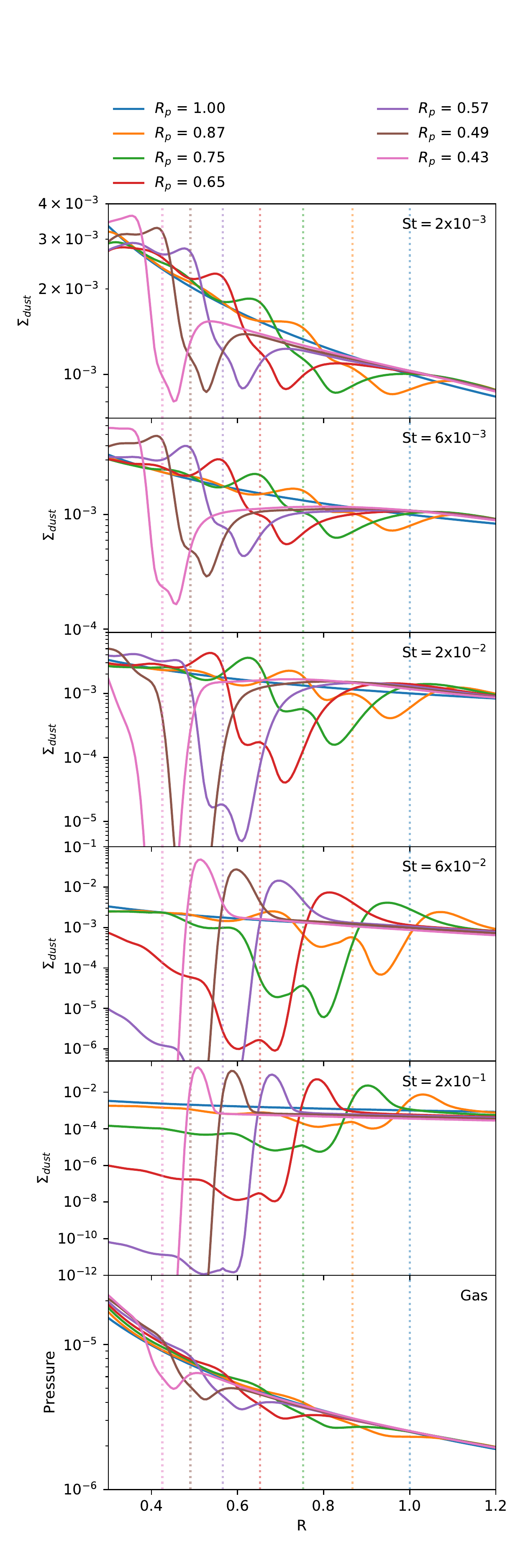}
\vspace{-1.0cm}
\caption{As with \autoref{fig:12MEarth_timeevol} but for a $20 \MEarth$ planet.}
\label{fig:20MEarth_timeevol}
\end{figure}

\begin{figure}
\vspace{-1.75cm}
\includegraphics[width=1.0\columnwidth]{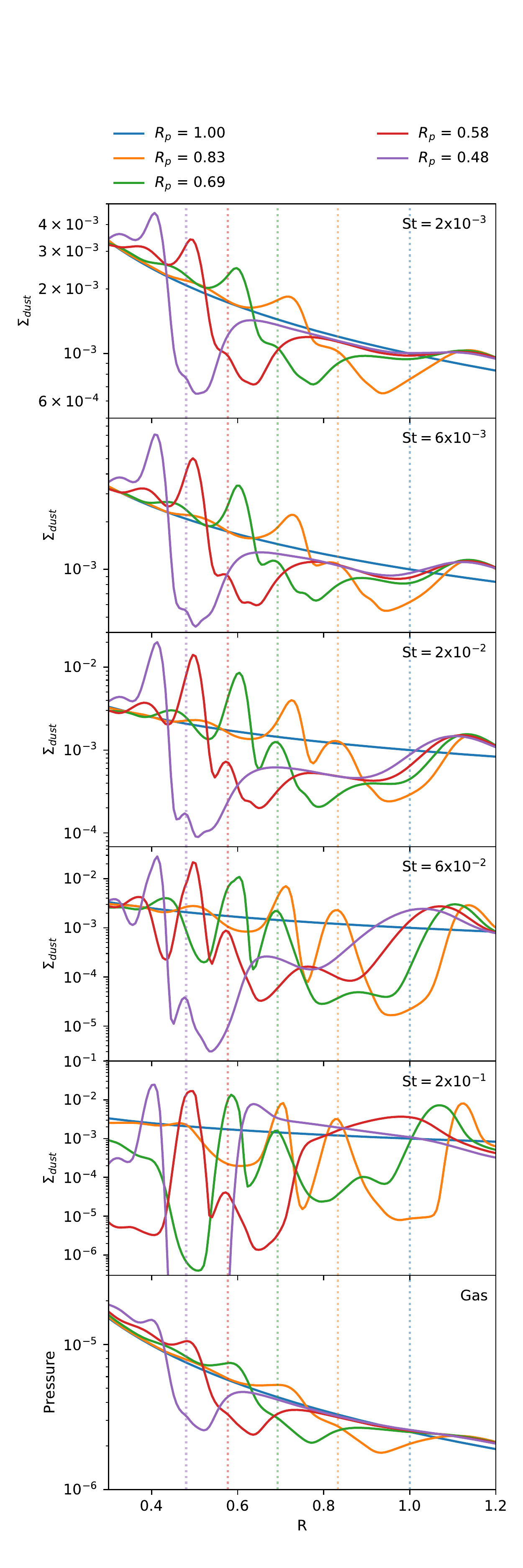}
\vspace{-1.0cm}
\caption{As with \autoref{fig:12MEarth_timeevol} but for a $60 \MEarth$ planet.}
\label{fig:60MEarth_timeevol}
\end{figure}

\autoref{fig:12MEarth_timeevol}, \ref{fig:20MEarth_timeevol} and~\ref{fig:60MEarth_timeevol} show that the trend of interior rings for small dust sizes and exterior rings for larger dust sizes, observed for $30 \MEarth$ planets, is present here for 12, 20 and $60 \MEarth$ planets.  The switch from a dominant interior dust ring to a dominant outer dust ring occurs at lower Stokes numbers for lower mass planets.  This is because the Type I migration rate is slower for low mass planets and so the smaller dust in the outer disc is more likely to keep up with the pressure perturbation (maximum or point of inflection).  We also note that for the $60 \MEarth$ simulation where an additional pressure perturbation forms exterior to the planet, the dust enhancement in the outer disc for large particles primarily occurs closer to the outer bump (near the planet's original location).

\subsection{Outer bump gap profile for high mass planets}
\label{sec:60MEarth_gap}

There is a complex gap profile (for the highest mass planets) associated with the 'outer bump' in gas surface density described in Appendix~\ref{sec:gas}.  We argue in Section~\ref{sec:GasAndToy} that this phenomenon is associated with a transient accumulation of gas  outside the planet's initial location which then diffuses away on a viscous timescale. We see in \autoref{fig:60MEarth_timeevol} (for a $60 \MEarth$ planet) that for the lowest Stokes numbers, the dust density maximum exterior to the planet is found near the outer bump. For the highest Stokes number shown, where the dust exterior to the planet is migrating faster than the planet, the dust density maximum eventually shifts from the vicinity of the outer bump to the pressure maximum exterior to the planet (purple contour for $\St = 0.2$).  It is also possible for intermediate Stokes numbers (see purple and red contours for $\St = 2 \times 10^{-2}$ and $6 \times 10^{-2}$) that there can temporarily be {\it two} dust density maxima exterior to the planet, associated respectively with the pressure maximum just outside the planet and with the 'outer bump' outside the planet's original location. This sounds a note of caution about interpreting multiple rings as requiring multiple planets. Whether this phenomenon of multiple exterior rings from a single planet would ever be observable (and what errors in planet mass might be introduced by misidentifying the ring at the outer bump as representing the location of the pressure maximum) is a matter for future investigation.  It is important to note that even in the case of two dust density maxima exterior to the planet, the interior dust ring dominates as it has a stronger density peak.  Furthermore the additional pressure perturbation in the outer disc can trap large sized dust, thus starving the region interior to this of large dust grains and potentially suppressing the dust collection in the pressure perturbation that is exterior to and moves with the planet.  This may then cause the planet location to be further misdiagnosed.

\section{Resolution test}
\label{sec:appendixC}

\begin{figure}
\vspace{-1.75cm}
\includegraphics[width=1.0\columnwidth]{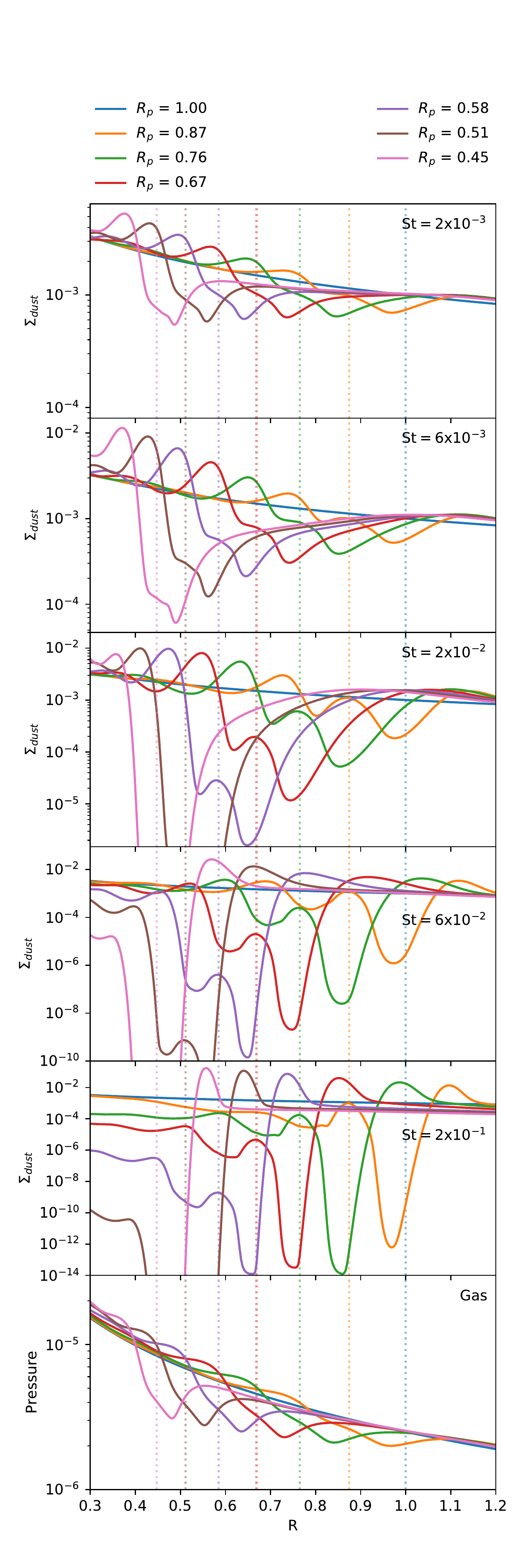}
\vspace{-1.0cm}
\caption{Time evolution of the dust density profile for particles of various Stokes numbers (top five panels) and time evolution of the pressure profile (bottom panel) for the $30 \MEarth$ migrating planet simulation performed at double the resolution (equivalent to Figure~\ref{fig:30MEarth_timeevol}).  The key results, i.e. that the small- and large-sized dust form a ring interior and exterior to the planet, respectively, are unaffected by resolution.}
\label{fig:30MEarth_timeevol_HR}
\end{figure}

\begin{figure}
\vspace{-0.7cm}
\includegraphics[width=1.0\columnwidth]{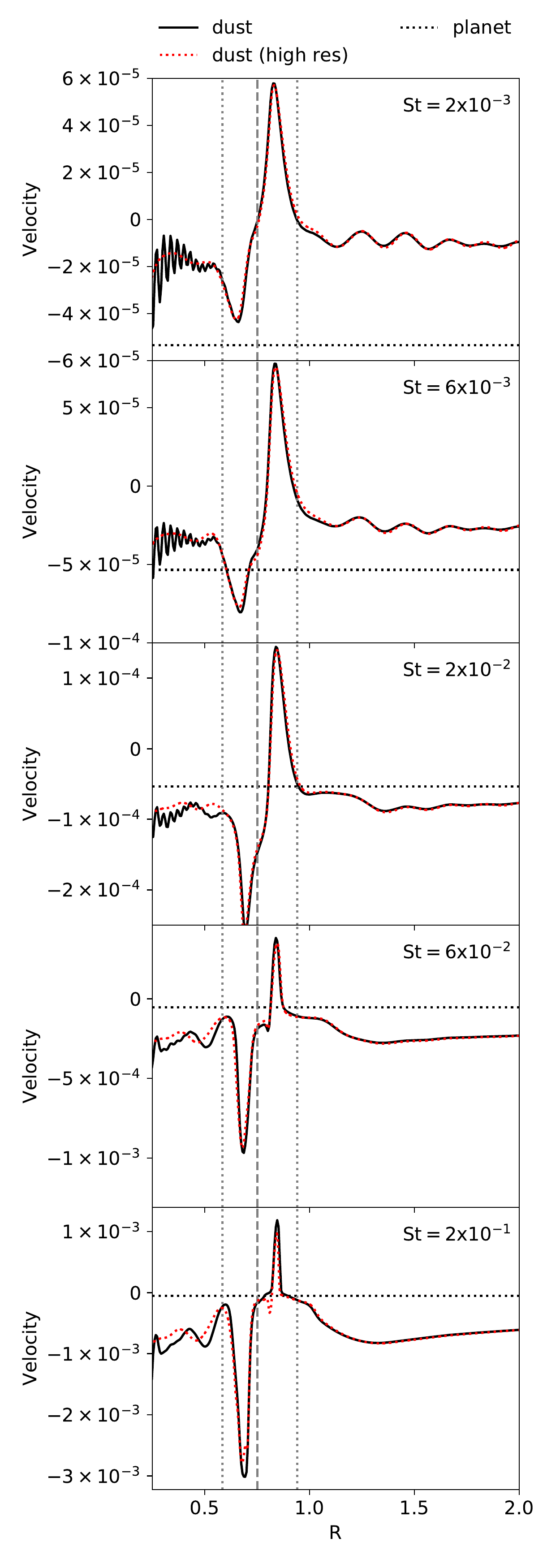}
\vspace{-0.8cm}
\caption{Dust velocity against disc radius compared to the instantaneous planet velocity (horizontal dotted line) when the planet is at $\Rp = 0.75$ for various Stokes numbers for the simulation presented in \autoref{fig:vdust_vp} (solid black) as well as the high resolution simulation (red dotted).  The velocities are approximately the same but note the removal of oscillations in the inner disc for low Stokes numbers at the higher resolution.  The locations of the planet and the interior and exterior pressure perturbations are marked with vertical dashed and dotted grey lines, respectively.} 
\label{fig:vdust_vp_HR}
\end{figure}

We perform a resolution test using twice as many grid cells in both the radial and azimuthal directions for the $30 \MEarth$ planet simulation.  \autoref{fig:30MEarth_timeevol_HR} shows the time evolution of the dust density profile -- equivalent to \autoref{fig:30MEarth_timeevol}.  The small dust produces an interior ring while the large dust forms an exterior ring, with the transition occurring between ${\rm St} \approx 2 \times 10^{-2}$ and $6 \times 10^{-2}$, just as in the low resolution simulations.  \autoref{fig:vdust_vp_HR} shows the dust velocity against radius and is the equivalent to \autoref{fig:vdust_vp} but with the high resolution simulation results overplotted.  It is clear that the results are similar to \autoref{fig:vdust_vp} and that the transient oscillations for small Stokes numbers disappear at higher resolution.  Given that the overall result is unaffected (\autoref{fig:30MEarth_timeevol_HR}) these oscillations do not impact the main conclusion of this paper.

\section{Comparison of back-reaction and migration timescales}
\label{sec:appendixB}

\begin{figure}
\includegraphics[width=1.0\columnwidth]{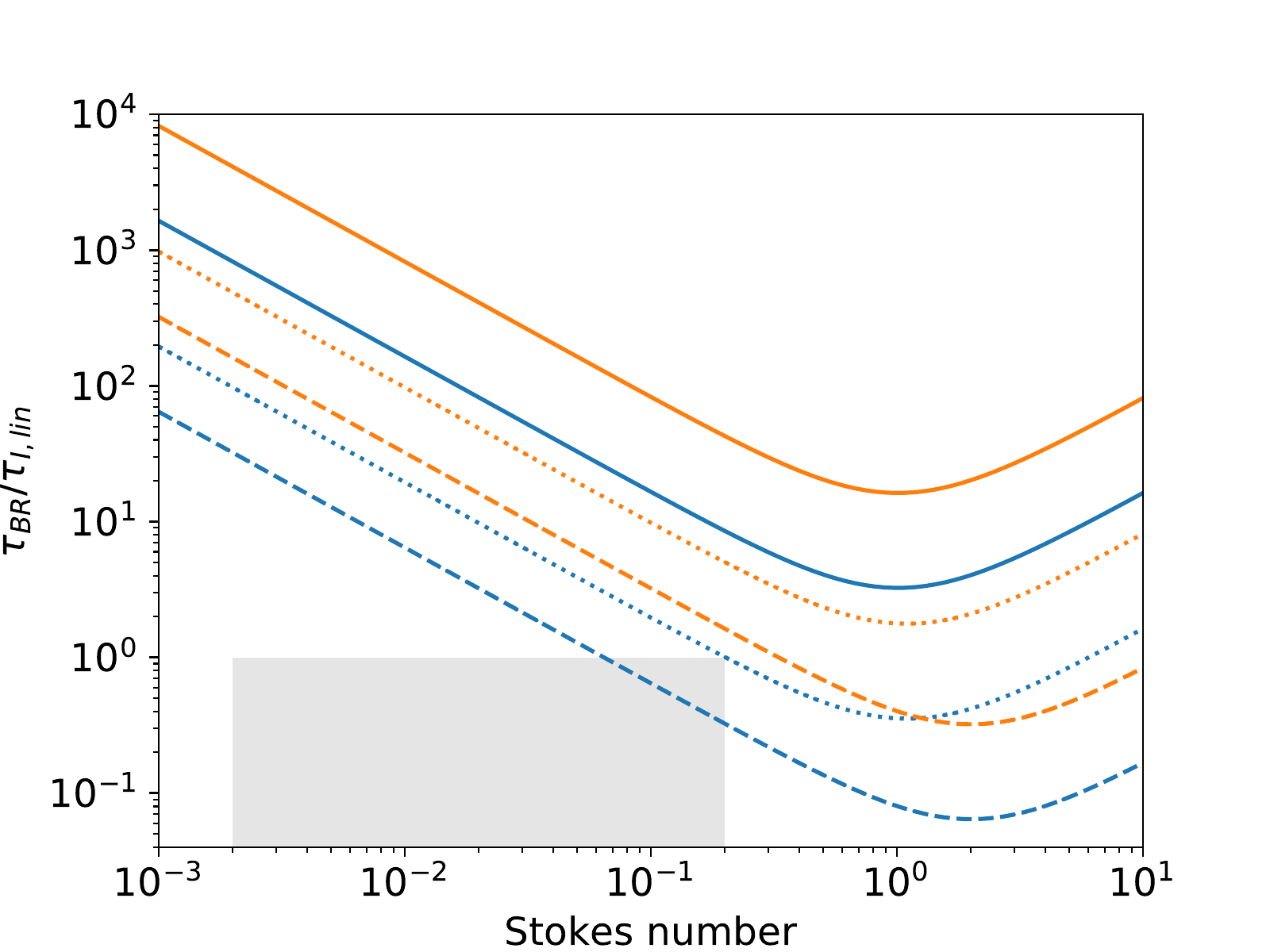}
\caption{Ratio of the back-reaction timescale to the migration timescale for the disc setup used in this study for the 12 (blue) and 60 $\MEarth$ (orange) planets and for dust-to-gas ratios of 0.01 (solid), 0.1 (dotted) and 1 (dashed).  Back-reaction becomes important when this ratio is less than unity.  The grey solid region shows the regime where it is important for the Stokes numbers explored in this paper.  Back-reaction is only important for the low mass planets for large Stokes numbers if the dust-to-gas ratio reaches unity.}
\label{fig:BR}
\end{figure}

Using the notation in Equation C1 of \cite{Gonzalez_BR}, the velocity of the gas associated with the back-reaction term is given by

\begin{equation}
v_{\rm BR} = \frac{\Delta R}{\tau_{\rm BR}} = - \frac{f_{\rm drag}}{\Sigma_{\rm g} \Omega} \pderiv{}{R} (c_{\rm s}^2 \Sigma_{\rm g}),
\label{eq:v_BR}
\end{equation}
where $\tau_{\rm BR}$ is the back-reaction timescale and

\begin{equation}
f_{\rm drag} = \frac{\epsilon}{(1+\epsilon)^2 \St^{-1} + St},
\label{eq:f_drag}
\end{equation}
where $\epsilon = \rho_{\rm d}/\rho_{\rm g}$ is the dust to gas ratio.

We use the analytical expressions for the back-reaction and the migration timescales (equations~\ref{eq:v_BR} and~\ref{eq:f_drag} and~\ref{eq:tauI_lin})  to determine the regime in which back-reaction becomes important for our disc setup.  Figure~\ref{fig:BR} shows the timescale on which the back-reaction occurs compared to the migration timescale against Stokes number for the 12 and 60 $\MEarth$ planets (i.e. spanning the planet mass regime that we study), for various different dust-to-gas ratios.  It is clear that back-reaction only starts to become important for $\St > 0.1$ and for high dust-to-gas ratios.  Thus for typical protoplanetary discs where the dust-to-gas ratio is expected to be small, we do not expect back-reaction to be important for the regime we explore.


\bsp	
\label{lastpage}
\end{document}